\documentclass[11pt,a4paper]{article}
\usepackage{multirow}
\usepackage{subfig}
\usepackage{graphicx}
\usepackage{epsfig}
\usepackage{braket}
\usepackage{jheppub}

\def  \bcen   {\begin{center}}
\def  \ecen   {\end{center}}

\newcommand{\beq}{\begin{eqnarray}}
\newcommand{\eeq}{\end{eqnarray}}

\newcommand{\centeron}[2]{{\setbox0=\hbox{#1}\setbox1=\hbox{#2}\ifdim                                       
\wd1>\wd0\kern.5\wd1\kern-.5\wd0\fi \copy0
\kern-.5\wd0\kern-.5\wd1\copy1\ifdim\wd0>\wd1 \kern.5\wd0\kern-.5\wd1\fi}}
\newcommand{\ltap}{\>\centeron{\raise.35ex\hbox{$<$}}
                               {\lower.65ex\hbox{$\sim$}}\>}
\newcommand{\gtap}{\>\centeron{\raise.35ex\hbox{$>$}}
                               {\lower.65ex\hbox{$\sim$}}\>}

\newcommand\ZZ{\hbox{\zfont Z\kern-.4emZ}}
\font\zfont = cmss10 

\newcommand{\X}{$X_{5/3}$ }
\newcommand{\Xbar}{$\overline{X}_{5/3}$ }

\title{Heavy vector-like quark with charge 5/3 at the LHC}    

\author[a]{Giacomo Cacciapaglia,}
\author[a]{Aldo Deandrea,}
\author[b]{Luca Panizzi,}
\author[a]{St\'ephane Perries}
\author[a]{and Viola Sordini}
\affiliation[a]{Universit\'e de Lyon, France; Universit\'e Lyon 1, CNRS/IN2P3, UMR5822 IPNL,\\ F-69622 Villeurbanne Cedex, France}
\affiliation[b]{School of Physics and Astronomy, University of Southampton,\\ Highfield, Southampton SO17 1BJ, UK.}

\emailAdd{g.cacciapaglia@ipnl.in2p3.fr}
\emailAdd{deandrea@ipnl.in2p3.fr}
\emailAdd{s.perries@ipnl.in2p3.fr}
\emailAdd{sordini@ipnl.in2p3.fr}
\emailAdd{l.panizzi@soton.ac.uk}

\abstract{We study the phenomenology at the Large Hadron Collider of an exotic vector-like quark with charge +5/3. We relax the
assumption of  a 100\% branching into $W t$ and allow for an arbitrary rate into $W$ plus light quarks, thus covering all possible 
scenarios. Sizeable decays into light quarks can be achieved, for instance, in a model where the \X quark is embedded in a doublet with 
hypercharge 7/6, which also contains a $t'$ quark. We study the bounds on the parameter space of this model, and perform a detailed 
simulation of the \X pair production and decays. We show that the final state with $W t W q$, where $q=u$ or $c$, contributes to the 
same sign dilepton searches, and that the reach can be improved with alternative cuts optimised on such final state.}


\preprint{LYCEN 2012-02} 

\makeindex

\begin{document}

\maketitle
\flushbottom

\tableofcontents

\section{Introduction}

The top quark certainly plays a special role in the Standard Model (SM), because it is the only fermion with a mass close to the 
electroweak scale, thus coupling strongly with the Higgs sector. Any model of new physics which attempts to solve the hierarchy 
problem, or that aims at modifying the properties of the Higgs boson, will likely predict the presence of heavy partners of the top.
Among these new states, vector-like quarks occupy a special place for their role in model building, as they are often predicted by 
models based on new symmetries: in extra dimensional models, for example, they appear as Kaluza-Klein recurrences of the 
standard model quarks; in little Higgs type models they ensure the cancellation of the top Yukawa divergent loop~\cite{LHreview}; 
in models where a new strongly-coupled sector is responsible for the breaking of the electroweak (EW) symmetry they emerge 
as part of the composite sector~\cite{GHUwarped} or are predicted when a custodial symmetry 
is introduced to protect the coupling $Zb\bar b$ from large deviations \cite{Agashe:2006at}.
The presence of TeV scale vector-like quarks is also an interesting possibility for their effect on the Higgs properties: for instance, 
supersymmetric models with vector-like quarks can ease the MSSM tension in the Higgs mass whose value is lifted from the tree level 
prediction $m_H \sim m_Z$ via loop contributions~\cite{Moroi:1991mg,Martin:2009bg,Graham:2009gy}; they have also been proposed as a 
solution~\cite{Azatov:2012rj, Bonne:2012im,Moreau:2012da} for the apparent excess in the $H \to \gamma \gamma$ rate 
observed by both CMS and ATLAS (for a general discussion of the effect of new particles to the $H \to \gamma \gamma$ decay mode
see \cite{Cacciapaglia:2009ky,Cacciapaglia:2012wb}).
Vector-like quarks are also phenomenologically interesting as, depending on the choice of their quantum numbers, they may be 
much less constrained than a fourth family of standard model like quarks. 
By simply requiring the presence of a Yukawa-type coupling with the standard quarks, the allowed representations can contain both 
a top and bottom partner ($t'$ and $b'$), but also two exotic quarks with charges $+\frac{5}{3}$ ($X_{5/3}$) and 
$- \frac{4}{3}$ ($Y_{-4/3}$)~\footnote{More exotic charges, like $+8/3$ and $-7/3$, can be obtained from larger SU(2) representations when more than one vector-like quark are included.}. In~\cite{Cacciapaglia:2010vn} a complete list of the possible representations is presented, including the 
relevant bounds from precisely measured quantities in the Standard Model (SM).
For earlier bounds see for example \cite{Alwall:2006bx,delAguila:2008iz}. However the most competitive bounds will come in the 
short term from direct searches at the Large Hadron Collider (LHC). The LHC is indeed collecting data that allow to test this 
sector quite precisely up to a few hundred GeV, and many bounds have already been published based on the 2011 dataset at 7 TeV, 
while the 2012 run at 8 TeV awaits to be analysed. A review of the 2011 bounds can be found in~\cite{Okada:2012gy}. 
However precise bounds crucially depend on the decay modes of the new states.
In the experimental searches, the simplified assumption is often to consider 100\% branching ratio in a single channel, and pair 
production via QCD interactions, which is independent on the charge and electroweak couplings of the new quark.
However, no realistic case corresponds to this simplified assumption, and the bounds often cannot be translated in a straightforward 
manner to more realistic cases.
In fact, the presence of various decay modes means that channels with mixed decays will also contribute to the signal region with an 
unknown experimental efficiency: a simple rescaling of the cross section with the branching ratios, therefore, is not sufficient to extract 
a reliable bound. Furthermore, typical cross-sections for single production of heavy vector-like 
quarks can become competitive with QCD pair production at high mass, even if they are model dependent.
The exclusion limits from the 2011 data already fall in the mass region where both production channels are fairly large and may give 
comparable contribution to the signal. 

Vector-like quarks have recently been the subject of
wide interest as they may be the first hint of physics beyond the standard model to be seen at the LHC (for a general review 
see for example \cite{Okada:2012gy,AguilarSaavedra:2009es}).
Heavy vector-like quarks typically decay to SM quarks plus $W$ and/or $Z$ and Higgs bosons. 
Several decay modes of the $t'$ and $b'$ have been considered both by experimental searches and phenomenological studies, and a 
dedicated search for the exotic one $X_{5/3}$ is available from CMS~\cite{CMSX53}.
In this paper we will focus in detail on the production and decays of such a state with electric charge $5/3$, which is motivated, for example, 
by a custodial symmetry in models of composite Higgs~\cite{Agashe:2006at}.
It can only interact with SM up-type quarks through charged currents, while decays to a $t'$ are only allowed for large mass splitting.
Due to the exotic charge, its main distinctive decay mode is $X_{5/3} \to W^+ t \to W^+ W^+ b$, which can give rise to two same-sign 
leptons \cite{Contino:2008hi} via the $W$ leptonic decays. This peculiar state was for example used in
composite Higgs models searches \cite{Mrazek:2009yu,Dissertori:2010ug}.
The same sign dilepton signature can also arise from the same final state with 4 $W$'s and 2 $b$'s which is also produced in the case 
of a $b'$ quark decaying into $b' \to W^- t \to W^- W^+ b$, therefore, for example, the bound on the $b'$ 
mass in~\cite{Chatrchyan:2012yea,Aad:2012bb} can be directly translated into a bound on the \X mass because the experimental 
efficiencies in the two cases are very similar~\cite{Matsedonskyi:2012ym}. 
Both phenomenological studies and the dedicated CMS search consider 100\% branching ratio into the top and pair production, while, 
in realistic set-ups, mixing with light fermions $c$ and $u$ can allow decays into $W$ and a light jet.
Such mixing is tightly bound by flavour observables, however depending on the case, the bounds can be fairly 
mild~\cite{Cacciapaglia:2011fx} and the branching ratio into light quarks can be competitive with the top one.
Furthermore, the decay into light quarks may imply the presence of sizeable single \X production mediated by the same coupling 
with the $W$ boson. On general ground, therefore, one cannot ignore the possibility of a sizeable coupling to light quarks.
In light of this observation, in this paper we propose a more detailed study, where we consider pair production of the \X quark with more 
complete and realistic decay modes, allowing for decays into light quarks.
The channel where both \X decay into $W$ jet can be constrained by a search for heavy quarks decaying into $W q$, with $q$ being 
any light quark~\cite{Aad:2012bt}. In this paper we do not consider single production of the \X state because it is a more 
model-dependent channel as it depends on the precise value of the mixing.
We therefore postpone the single production analysis to a dedicated study, where we shall also address the matching procedure for the 
inclusion of QCD radiation, which is well under control only in the pair production case at present. 

The results of our study can be applied to any model containing an exotic \X quark, as the only relevant parameters are the mass of 
the quark (which determines the pair production cross section) and the branching ratio into light quarks.
However, to be more concrete and without loss of generality, our simulation is based on a model where the \X is embedded in a 
non-SM doublet with hypercharge $7/6$, which also contains a $t'$~\cite{Cacciapaglia:2011fx}.

\section{Mixing parameters in the non-SM doublet model}

The phenomenology of the \X quark in the pair production channel can be fully characterised in terms of two free parameters:
\begin{itemize}
\item[-] the mass $M_X$, which will determine the QCD production cross section;
\item[-] the branching ratio into tops, $BR (X_{5/3} \to W^+ t) = b$.
\end{itemize}
The branching into light quarks is therefore given by
$$
BR (X_{5/3} \to W^+ u, W^+ c) = 1-BR (X_{5/3} \to W^+ t) = 1-b\,.
$$
We are assuming here that \X cannot decay into $W^+ t'$ because such mode is not kinematically accessible. This assumption is quite 
general because in any reasonable model the splitting between the \X and $t'$ belonging to the same SU(2) multiplet can only be 
generated by Yukawa-type interactions via the Higgs vacuum expectation value (VEV), and such contribution is limited in size.

Without loss of generality, we will base our simulations on a specific simple model containing a vector-like quark which is an $SU(2)$ 
doublet with hypercharge $7/6$ (corresponding to a shift of one hypercharge unit with respect to SM quark doublets). 
Such a doublet also contains a top partner $t'$ which is studied in the literature (see for example  \cite{Cacciapaglia:2011fx}
and references therein) due to its mixing with SM up-type quarks. The mass eigenstates are therefore:
\begin{eqnarray}
\setlength{\arraycolsep}{0pt}
\begin{array}{cccccccccccc}
&&&&&&&&&\begin{multirow}{2}{*}{\Big(}\end{multirow}&X_{5/3}&\begin{multirow}{2}{*}{\Big)}\end{multirow}\\
\begin{multirow}{2}{*}{\Big(}\end{multirow}&u&\begin{multirow}{2}{*}{\Big)}\end{multirow}&
\begin{multirow}{2}{*}{\Big(}\end{multirow}&c&\begin{multirow}{2}{*}{\Big)}\end{multirow}&
\begin{multirow}{2}{*}{\Big(}\end{multirow}&t&\begin{multirow}{2}{*}{\Big)}\end{multirow}&&t^\prime\\
&d&&&s&&&b&
\end{array}
\end{eqnarray}
In general, the $t^\prime$ mixes with all up-type quarks, and details of the mixing matrices for the general case of mixing with 
all three standard model families can be found for example in \cite{Cacciapaglia:2011fx}. 
In this section we will simply recall the results relevant for our discussion of the \X phenomenology.
 
If the SM Higgs sector is not extended by the introduction of new scalars, the new doublet can only couple with the Higgs and the SM 
right-handed up-type singlets. The new Yukawa couplings generates a mixing in the up sector, with the three lighter mass eigenstates 
identified with the SM quarks.  Denoting by $u^i_R$ and $u^i_L$ the SM singlets and up component of the doublets, and by $( X, U)^T$ 
the new doublet, the resulting mass terms are as follows (in the basis where the SM Yukawa couplings $y_u$ are diagonal):
\beq
\mathcal{L}_{\rm mass} = - \sum_{i=1}^3\, \frac{y^i_u v}{\sqrt{2}}\, \bar u^i_L u^i_R -  \sum_{i=1}^3\, x_i\, \bar U_L u^i_R - M\, \bar U_L U_R  - M\, \bar X_L X_R  + h.c.\, ,
\eeq
where $v \sim 246$ GeV is the Higgs VEV, $y_u^i$ are the diagonal SM up Yukawas, $M$ is the vector-like mass (also equal to the 
mass of $X_{5/3}$, $M_{X}=M$) and $x_i$ encode the mixing generated by the Higgs VEV ($x_i=\frac{\lambda^i v}{\sqrt{2}}$, where 
$\lambda^i$ are the new exotic Yukawa couplings).
In the down sector, only the SM Yukawa matrix remains.
We can first diagonalise the mass matrix for top and $t'$, thus neglecting the mixing with the light quarks:
\begin{eqnarray}
\left(
\begin{array}{cc}
\cos \theta^{L}_{u} & -\sin \theta^{L}_{u} \\
\sin \theta^{L}_{u} & \cos \theta^{L}_{u} \\
\end{array}
\right)
\left(
\begin{array}{cc}
\frac{y_t v}{\sqrt{2}} & 0 \\
x_3 & M \\
\end{array}
\right)
\left(
\begin{array}{cc}
\cos \theta^{R}_{u} & \sin \theta^{R}_{u} \\
-\sin \theta^{R}_{u} & \cos \theta^{R}_{u} \\
\end{array}
\right) =
\left(
\begin{array}{cc}
m_{t} & 0 \\
0 & m_{t^{\prime}} \\
\end{array}
\right)\, ;
\label{masseigenstates}
\end{eqnarray}
where:
\beq
\sin \theta^{R}_{u} = \frac{M x_3}{\sqrt{(M^{2} - m_{t}^{2})^{2} + M^{2} |x_3|^{2}}}\, , \qquad \sin \theta^{L}_{u} = 
\frac{m_t}{M} \sin \theta^{R}_{u}\, .
\eeq
Note that the mixing angle for left-handed components $\theta_L$ is typically smaller that the right-handed one $\theta_R$, being 
suppressed by the mass of the top over the heavy mass $M$ (which is typically larger).
This property is still valid in the case of mixing with three flavours, with the left-handed mixing terms proportional to the light quark 
masses, while the right-handed ones are proportional to the Yukawa masses $x_i$. 
The mixing matrices are replaced by $4\times 4$ unitary matrices $V_L$ and $V_R$, acting respectively on the left- and right-handed 
components. The mass of $t'$ is given by
\beq
m_{t'}^2 = M^2 \left( 1 + \frac{|x_3|^2}{M^2 - m_t^2} \right) \left( 1 +  \frac{(|x_1|^2 + |x_2|^2) \cos^2 \theta_R}{M^2} + \dots\right) \,, \label{eq:masstp}
\eeq
where we have kept the leading terms in $x_1$, $x_2$ and the light quark masses.
From this formula it is clear that the presence of the new Yukawas always increases the $t'$ mass so that $m_{t'} > M = M_X$, and that 
the splitting is typically small being suppressed by the vector-like mass $M$.
In this model, therefore, the decay mode $X_{5/3} \to W^+ t'$ is absent, and we can expect this conclusion to hold in general because of 
the small splitting between the two masses.

The couplings of \X with top, charm and up are generated by the coupling with its SU(2) partner $U$, which mixes with the SM up 
quarks. The couplings with the mass eigenstates read:
\beq
\mathcal{L}_{X} &=& i \frac{g}{\sqrt{2}}\, W^+_\mu\; \bar{X}_R \gamma^\mu \left( V_R^{44} t'_R + V_R^{43} t_R + V_R^{42} c_R + V_R^{41} u_R \right) + \nonumber \\
  & & i \frac{g}{\sqrt{2}}\, W^+_\mu\; \bar{X}_L \gamma^\mu \left( V_L^{44} t'_L + V_L^{43} t_L + V_L^{42} c_L + V_L^{41} u_L\right) + h.c.\,,
\eeq
where the relevant matrix elements are related to the new Yukawa couplings by:
\beq
\begin{array}{cccc}
V_R^{41} = - \frac{x_1}{M}\,, & V_R^{42} = - \frac{x_2}{M}\,, & V_R^{43} = - \sin \theta_R\,, & V_R^{44} = \cos \theta_R\,; \\
V_L^{41} = - \frac{m_u x_1}{M^2}\,, & V_L^{42} = - \frac{m_c x_2}{M^2}\,, & V_L^{43} = - \frac{m_t}{M} \sin \theta_R\,, & V_L^{44} = \frac{m_{t'}}{M} \cos \theta_R\,;
\end{array}
\eeq
while the remaining terms can be found in~\cite{Cacciapaglia:2011fx}.
The formulas show that the left-handed mixing with the light quarks, $V_L^{41}$ and $V_L^{42}$, can be safely neglected being 
suppressed by the light quark masses.
\begin{table}[tb]
\begin{center}
\begin{tabular}{|l|c|}
\hline
$D_0$--$\bar{D}_0$ mixing & $|V_R^{41}||V_R^{42}| < 3.2 \times 10^{-4}$ \\
\hline
APV in Cesium & $|V_R^{41}| < 7.8 \times 10^{-2}$ \\
\hline
LEP1, charm couplings & $|V_R^{42}| < 0.2$ \\
\hline
CMS: $t \to Z c, Z u$ & $|V_R^{43}| \sqrt{|V_R^{41}|^2 + |V_R^{42}|^2} < 0.07 |V_{tb}|$\\
\hline
\end{tabular}
\caption{Bounds on $|V_R^{41}|$ and $|V_R^{42}|$ from \cite{Cacciapaglia:2011fx}, and updated with the results in \cite{Chatrchyan:2012sd} for the top decays.} 
\label{tab:sumbounds}
\end{center}
\end{table}
The matrices $V_L$ and $V_R$ will also introduce a flavour mixing in the up sector: the most dangerous effect is the presence of flavour 
changing couplings to the $Z$ boson, due to the fact that the couplings of $U$ to the $Z$ differ from the couplings of the SM up quarks.
The down quark sector is not directly affected by the presence of the vector-like doublet,  
thus the precise measurements in the Kaon and B meson sector do not pose tight bounds in our case. 
The tree level FCNCs generated by the $Z$ couplings also depend on the same matrix elements appearing in the \X couplings, thus we 
can derive bounds on the couplings from flavour observables.
The most severe bounds come from the $D$ mesons, in particular from $D_0$--$\bar{D}_0$ mixing which is the most 
stringent bound on flavour violation for this non-standard doublet case. 
In particular the most important constraint are on the right mixing matrix $V_R$.
A summary of the relevant bounds on the mixing with the up $|V_R^{41}|$ and down $|V_R^{42}|$ can be found 
in \cite{Cacciapaglia:2011fx} and is summarised in Table~\ref{tab:sumbounds}. The bound on rare top decays has been updated with 
respect to the one quoted in \cite{Cacciapaglia:2011fx}: CMS has undertaken a study of rare top decay with an integrated luminosity of 
$5~\mathrm{fb}^{-1}$ \cite{Chatrchyan:2012sd}, finding a limit $\Gamma(t\to Zq)/\Gamma(t\to Wb)<0.24\%$.
The strong limit from $D_0$--$\bar{D}_0$ mixing is on the product of the mixing with the first and second generation, while the bounds 
on the individual mixing terms are rather mild, coming from Atomic Parity Violation (APV) experiments for the up and from LEP1 
measurements of the charm couplings to the $Z$ boson. While respecting the 
other bounds given in Table \ref{tab:sumbounds}, either one of the two mixings can be fairly large, thus giving significant branching ratios 
of  \X into $Wc$ and $Wu$. 
The branching depends on the mixing with the top, which is bound from electroweak precision measurements ($T$ parameter): the precise bound is mass dependent, however it is typically of the order $|V_R^{43}| = \sin \theta_R < 0.3$.
In the following section we further discuss this point.

\subsection{\X production and decay}
\label{sec:proddecay}

\begin{figure}[tb]
\centering
\epsfig{file=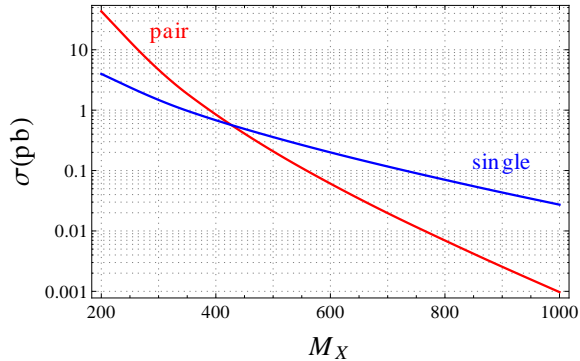, width=0.5\textwidth}
\caption{Production cross-sections of \X as function of the mass $M_X$. In blue the single production cross-section $pp\to X_{5/3}+Y$, 
where $Y=W^\pm$ or a light quark; in red the pair production contribution $pp\to X_{5/3}\bar X_{5/3}$. The mixing parameters, which 
determine the single production, are $\sin \theta_R = 0.3$, $V_R^{42}=0.2$, $V_R^{41} = 3.2\times10^{-4}/V_R^{42} = 1.6 \times 10^{-3}$.}
\label{fig:prodcross}
\end{figure}

\begin{figure}[tb]
\centering
\epsfig{file=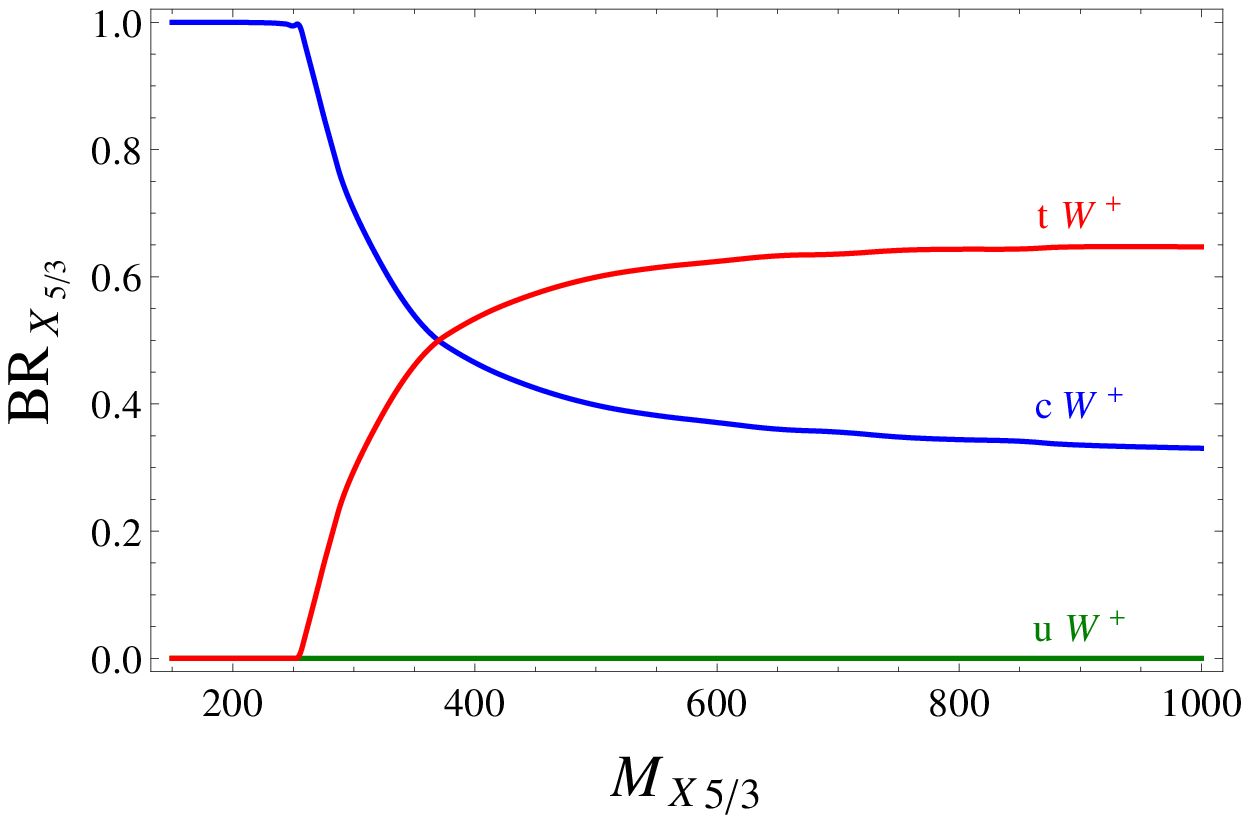, width=0.47\textwidth}\hfill
\epsfig{file=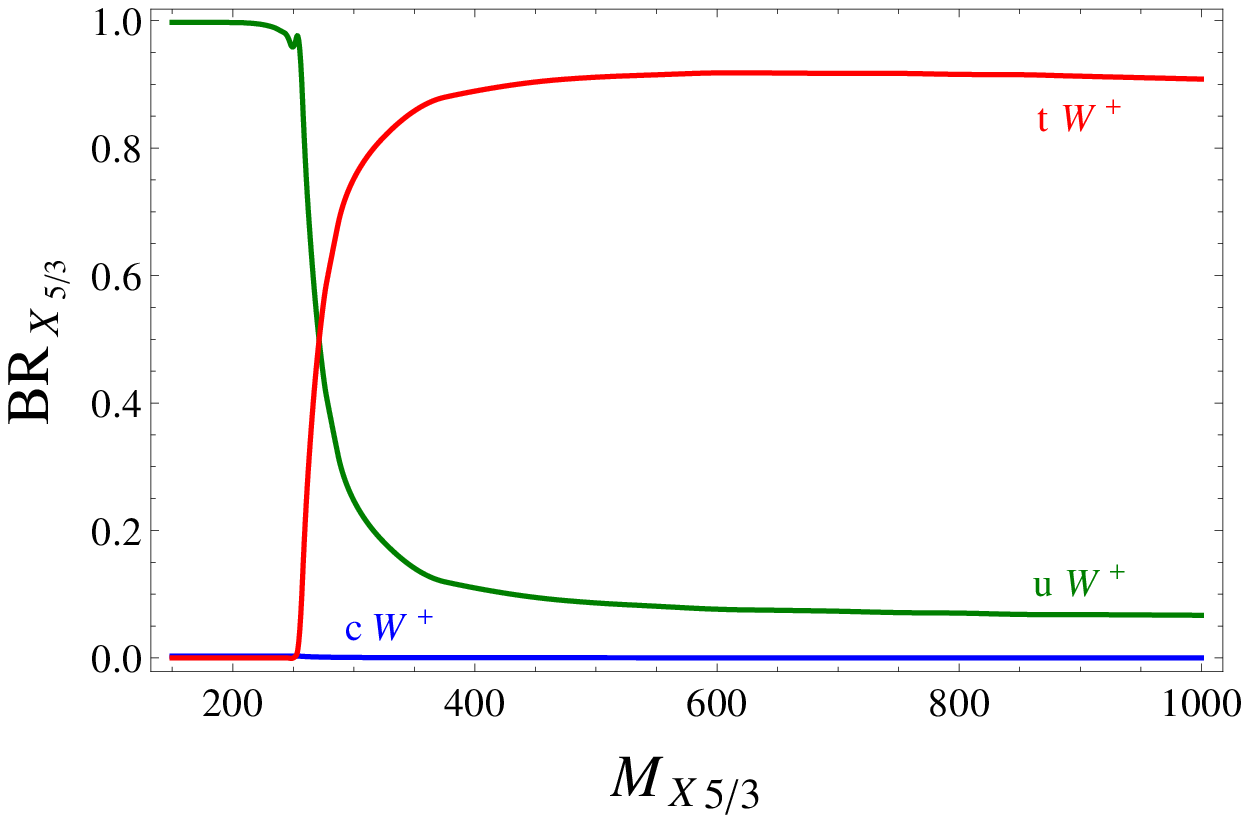, width=0.47\textwidth}
\caption{Branching ratios of \X as function of \X mass. In the left panel the mixing parameters are $\sin \theta_R = 0.3$, $V_R^{42}=0.2$, 
$V_R^{41} = 3.2\times10^{-4}/V_R^{42}$. This particular choice maximises $V_R^{42}$. In the right panel the mixing parameters are 
$\sin \theta_R = 0.3$, $V_R^{41}=0.078$, $V_R^{42} = 3.2\times10^{-4}/V_R^{41}$. This particular choice maximises $V_R^{41}$.}
\label{fig:BRX}
\end{figure}

The \X vector-like quark can be both single and pair produced. Pair production is dominated by QCD processes from gluon pairs and 
quark pairs. At LHC energies the production from gluons is dominant. The corresponding couplings are model independent as only 
dictated by the colour quantum numbers. Electroweak diagrams are sub-dominant, coming from quark pairs into gauge bosons. 
Single production is more model dependent because it occurs via quark 
pairs going into \X and standard quarks through the exchange of a $W$ boson. One can note that pair production, 
being mainly of QCD origin is dominant at low masses of the \X vector-like quark but decreases faster than single production when the 
mass of the \X quark increases. 
This is a well-known behaviour, which is due to the decrease of the parton distribution functions with the centre of mass energy of the 
parton-level collision. The detailed behaviour as a function of the \X quark mass is given in Figure~\ref{fig:prodcross} for a 
specific choice of the mixing parameters.
The \X vector-like quark can decay into up-type quarks ($X_{5/3}\to W^+ u_i$) but not into $W^+ t^\prime$ for the 
following reasons: 
\begin{itemize} 
\item if the \X and $t^\prime$ are lighter than the top quark, the mass gap between \X and $t^\prime$ is too small. This case is 
nevertheless disfavoured by present bounds on new heavy states;
\item if the \X and $t^\prime$ are heavier than the top quark, the mass hierarchy is $M_X<m_{t^\prime}$, as explained in the previous section. 
\end{itemize}
The branching ratios for parameters saturating the flavour bounds and for fixed $\sin \theta_R = 0.3$, 
are shown in Figure~\ref{fig:BRX}. 
It is possible to see that choosing to maximise the mixing with the charm, $|V_R^{42}|$, the 
decay into $W u$ is negligible: this is due to the constraint on the product of mixing parameters coming from the $D_0-\bar D_0$ 
mixing. On the other hand, choosing to maximise the mixing with the up, $|V_R^{41}|$,
the situation would be reversed, with the decay into $W c$ being irrelevant. 
The plots show in both cases sizeable decay rates into light quarks: this rate can be further enhanced if one chooses smaller mixing with 
the top, and can be 100\% in the extreme limit $\sin \theta_R = 0$.
As this possibility is not excluded, the branching into light quarks is virtually a free parameter.
The single production cross section, on the other hand, is proportional to the mixing parameters and therefore is crucially dependent on how large $V_R^{42}$ and $V_R^{41}$ are, 
while the production cross section for the \X pair is in practice insensitive to the particular choice of mixing in the light sector. \\
To summarize, when considering QCD pair production, the only dependence on the specific vector-like representation is encoded in the branching ratios, 
but this dependence can be completely removed if experimental efficiencies in the different channels were known. 
The phenomenology of different scenarios could then be described by simple rescalings of experimental data.
In the following we choose a benchmark point that maximises $|V_R^{42}|$: in this case the branching ratio to $W t$ is more 
suppressed, representing a scenario which is different from the one commonly searched in experimental analyses. We will then 
generalise our results in Section \ref{freeBR}, allowing the branching ratio to $W t$ to be a free parameter, in order to cover all possible parameter ranges and 
to allow a comparison to published experimental searches with $BR (X_{5/3} \to W t) = 100\%$.

\subsection{Present bounds}

In this section bounds coming from direct searches of heavy fermions at the LHC will be summarised. 
The most stringent direct bounds on the mass of the \X come from very recent searches for pair production of \X decaying 100\% 
into $Wt$ in both ATLAS and CMS experiments. However, processes involving down-type quarks decaying via $b^\prime \to Wt$, 
as the pair production $b^\prime {\bar b}^\prime \to W^- t W^+  {\bar t} $, have the same final state as the pair production of \X which decays 100\% to top quark and $W$ 
boson: $X_{5/3} {\bar X}_{5/3} \to W^+ t W^-  {\bar t} $, the only difference being that the charges of the two $W$s are flipped. 
Therefore, bounds coming from searches of pair production of $b^\prime \to W^-t$ can be directly translated in bounds for $X_{5/3} \to W^+t$.

The strongest bounds from ATLAS have been obtained in \cite{ATLAS-CONF-2012-130}, with a sample corresponding to an integrated luminosity of $4.7\mathrm{fb}^{-1}$: 
considering \X pair production and looking for a same-sign dilepton channel, the bound on the \X mass obtained by this search is $M_{X}>670$ GeV; the same search includes
also the contribution of single \X production under various assumptions on its couplling with $Wt$, and the obtained bound raises to $680-700$ GeV. 
In \cite{ATLAS-CONF-2012-137}, corresponding to an integrated luminosity of $4.64\mathrm{fb}^{-1}$, 
the single production of heavy quarks decaying only to light generations is studied: the quoted bound for \X is $m_X>1420$ GeV, but since
it relies on specific assumptions on the coupling between \X and light generations, it will not be considered in this study. 
ATLAS has also performed two analyses for a $b^\prime$ using $1.04\mathrm{fb}^{-1}$ \cite{Aad:2012bb, ATLAS:2012aw} finding that 
at 95\% confidence level a $b^\prime$ quark with mass $m_b^\prime < 480$ GeV decaying entirely via $b^\prime \to W t$ is excluded.

In CMS, the strongest bounds have been obtained in \cite{CMSX53} and \cite{Chatrchyan:2012af}. The first search has been performed in the same-sign dilepton channel using a 
data sample corresponding to an integrated luminosity of $5\mathrm{fb}^{-1}$, and the bound on the mass of \X obtained from that analysis is 
$M_{X}>645$ GeV. Since this search has been used to validate our analysis, more details will be illustrated in the following 
section. The second search \cite{Chatrchyan:2012af} was performed for a quark $Q$, 
pair produced by the strong interactions and decaying exclusively into a top quark and a W or Z boson. For a down type 
quark decaying 100\% into $Wt$, a mass below 675 GeV is excluded at the 95\% of confidence level. 
Previous direct bounds from CMS \cite{Chatrchyan:2012yea} on $b^\prime$ quarks obtained $m_b^\prime < 611$ GeV excluded at 
95\% confidence level assuming 100\% decay to top quark and $W$ boson. 

Note however that all these high mass bounds rely on the strong and unrealistic assumption that only one 
decay channel is present. In more realistic scenarios the \X quark can have sizeable decay modes to lighter quarks. 

Searches of $t^\prime$ with mixing only to third generation can also provide bounds for \X in our specific model, since $m_{t^\prime}$ is related to 
$M_X\,(=M)$ and to mixing parameters through the relation in Eq. (\ref{eq:masstp}).

The strongest bounds on $t^\prime$ by the ATLAS collaboration have been obtained in \cite{ATLAS:2012qe}, using a sample of $4.7\mathrm{fb}^{-1}$ and considering a lepton+jets final state: 
in this study the $t^\prime$ is supposed either to decay 100\% to $Wb$ or to decay both through charged and neutral currents to $Wb$,$Zt$ and $Ht$. Two bounds are therefore
provided: if the $t^\prime$ can only decay to $Wb$ it must be heavier than 656 GeV at 95\% C.L., while considering the more realistic scnearios with $BR(tW)=50\%$ and $BR(Ht)=BR(Zt)=33\%$, 
the bound is rescaled to 500 GeV. This search clearly shows how the bounds can be overestimated if considering just one decay channel.
In the single production search \cite{ATLAS-CONF-2012-137} decribed above, the bound for $t^\prime$ decaying 100\% to $Zq$ is 1080 GeV, but as already stated, this bound will not be considered
as it relies on specific assumptions.

CMS has searched for a $t^\prime$ which mixes with third generation. In \cite{Chatrchyan:2012vu}, a search for pair production of $t^\prime$ with $BR(Wb)$=100\% 
has been undertaken in the leptons+jet channel with a sample of $5\mathrm{fb}^{-1}$, finding the 95\% C.L. bound 
$m_{t^\prime}>560$ GeV; a similar search \cite{CMS:2012ab} at the same integrated luminosity obtains a nearly identical bound, 557 GeV. 
In \cite{Chatrchyan:2011ay}, using a sample of 
$1.14\mathrm{fb}^{-1}$, a search for pair production of $t^\prime$ assumed to decay 100\% into $tZ$ gives a bound on $t^\prime$ with 
mass $m_{t^\prime} > 475$ GeV at 95\% confidence level.

\section{Simulation Details}
Our study is performed for the LHC at 7 TeV in the centre of mass, assuming a luminosity of  $5\, \mathrm{fb}^{-1}$. 
The analysis is performed using {\tt MadGraph5} \cite{Alwall:2011uj} for generating the signal. Matching to {\tt Pythia} 
\cite{Sjostrand:2006za} is used for the hadronisation process, and a simplified detector simulation using {\tt Delphes} 
\cite{Ovyn:2009tx} to match the signal with the CMS experimental set up.
We chose the CMS set-up in order to validate our simulation with the published search in~\cite{CMSX53}, and to be able to 
compare our results.

\subsection{Signal}
The signal consists of inclusive pair production of \X plus QCD radiation and is given by the following process:
\begin{eqnarray}
 p p \to X_{5/3} \bar X_{5/3} (+ Y) + (n\leq2)~\mathrm{jets}
\end{eqnarray}
where $Y=W^\pm, Z, \gamma$ and the b-jets are included in the QCD radiation. Processes with additional electroweak gauge 
bosons in the final state are always suppressed by phase space and weak couplings (with the exception of $Y=\gamma$), but they 
have been kept nevertheless because they provide a source of leptons in the final state. The channels with $Y=t,t^\prime,h$ have a 
negligible cross section and will be ignored in the following analysis. The relative contributions of the different channels with the 
kinematic cuts described below and for different \X  masses are shown in Table \ref{channelweight}. We have quoted both LO and 
NLO+NNLL cross sections. The LO cross sections are those resulting from MadGraph before applying the matching procedure and 
are shown for reference purposes, while the NLO+NNLL cross sections, used to compare our results to CMS analyses, have been 
taken from Ref.\cite{CacciariNLO}. Note indeed that, after the matching, the contributions of the various sub-processes are distributed 
differently as they contain modified contributions from processes with jets.
The \X quark is always considered on-shell and it subsequently decays to $Wu_i$ with branching ratios which depend on the \X mass 
(see Figure~\ref{fig:BRX}). We have performed a scan over the \X mass in the range $\{300,700\}$~GeV with steps of 100~GeV. Below 
the $M_X=m_W+m_t\sim250$~GeV limit the decay into $Wt$ would not be allowed, strongly reducing same sign dilepton events in the 
final state.

The matching procedure has been performed following the $k_T$-MLM scheme, automatically implemented in {\tt Madgraph5+Pythia}, 
and the matching parameter \verb#xqcut# implemented in {\tt Madgraph5} has been computed in every step of the scan to optimise the 
matching procedure for each \X mass value. 
\begin{table}[tb]
\centering\begin{tabular}{l|cc|c|cc}
\begin{multirow}{2}{*}{$M_X$}\end{multirow} & \multicolumn{3}{c|}{cross sections [pb]} & 
\begin{multirow}{2}{*}{BR($Wt$)}\end{multirow} & 
Expected events \\
\cline{2-4}
 &  \multicolumn{2}{c|}{LO} & NLO + NNLL & & with $\ell^\pm \ell^\pm$ \\ 
(GeV)     &  $X_{5/3} \bar X_{5/3}$ &  $X_{5/3} \bar X_{5/3} + Y$ & $X_{5/3} \bar X_{5/3}$ & \% & \%  \\
\hline
 300 & 4.625 & 0.184 & 7.826 & 29.5 & 3.7\\
 400 & 0.847 & 0.032 & 1.379 & 53.4 & 6.7\\
 500 & 0.207 & 0.009 & 0.324 & 59.9 & 7.5\\
 600 & 0.061 & 0.003 & 0.091 & 62.4 & 7.8\\
 700 & 0.020 & 0.001 & 0.029 & 63.6 & 8.0 \\
\hline
\end{tabular}
\caption{Weights of different signal channels. Branching ratio of $X_{5/3}\to W t$ and expected rate of signal events with a pair of same sign leptons as function of the \X mass.}
\label{channelweight}
\end{table}

\subsection{Identification of SM backgrounds}
\label{subsec:smbg}
Reconstructing the \X quarks individually is a very difficult task, which suffers from huge combinatorial background.
In addition, in the complex environment of the LHC, final state objects overlap and their individual reconstruction is not efficient.
In order to get rid of the huge multi-jet background, we study the potential of a same sign dilepton signature. Depending on the \X mass 
and for our choice of parameters, this signal varies roughly from 2\% to 4\% of the total cross section (see Table~\ref{channelweight}).

In the following we consider only electrons and muons. 
There are two main categories of backgrounds: one is represented by $t\bar{t}$+jets, $V$qq (where $V$=$W$ or $Z$) and $Z$+jets. 
They do not contain opposite sign dileptons, but their cross-section is large enough to fake a same 
sign dilepton signature because of the charge misidentification of the leptons. 
The second category of background is the one with two real same sign dileptons. 
We have considered $VV$+jets, $Zt\bar{t}$+jets and $Wt\bar{t}$+jets.
The cross sections of the aforementioned processes are smaller, but they yield an irreducible background.
All these backgrounds and their cross-sections are summarised in Table \ref{tab:bckgdescription}.
Due to the recent discovery of a Higgs-like state at the LHC, we have also estimated the contribution of the $t \bar t H$ background, 
which falls within either the first or the second category depending on the decay channel of the Higgs boson. 
Assuming a Higgs mass of 125 GeV we have found that this background can be safely neglected, 
as will be shown in Section \ref{boundsfixedBR}.

\begin{table}[t!]
\begin{center}
\begin{tabular}{cc}
\hline
Background process & cross-section [pb]\\
\hline
$Z$+jets          &   2800 \\
$t\bar{t}$+jets &  165    \\
$Vqq$              & 36       \\ 
$VV$+jets       &   5        \\
$Wt\bar{t}$      & 0.169  \\
$Zt\bar{t}$       &  0.139 \\
\hline
\end{tabular}  
\caption{List of backgrounds considered in this analysis. The cross sections for all processes are at the LO, as 
obtained from the {\tt MadEvent} calculator~\protect{\cite{Alwall:2007st}}, except for $t\bar{t}$ events for which the approximate 
NLO calculation of~\protect{\cite{Kidonakis:2009mx}} is used. }
\label{tab:bckgdescription}
\end{center}
\end{table}

\subsection{Detector simulation}
\label{subsec:detsimu}
As explained before, background and signal events are generated with {\tt MadGraph/MadEvent}. In order to take into account the 
efficiency of event selection under semi-realistic experimental conditions, the detector simulation is carried out with the fast detector simulator 
{\tt Delphes}~\cite{Ovyn:2009tx} using the CMS detector model. The tracker is assumed to reconstruct tracks within $|\eta | < 2.4$ with 
a 100\% efficiency and the calorimeters cover a pseudo-rapidity region up to  $|\eta | <$ 3 with an electromagnetic and hadronic tower 
segmentation corresponding to the CMS detector. The energy of each quasi stable particle is summed up in the corresponding 
calorimeter tower. The resulting energy is then smeared according to resolution functions assigned to the electromagnetic calorimeter 
(EC) and the hadronic calorimeter (HC). The acceptance criteria are summarised in Table~\ref{Tab:DelphesAcceptance}. 
For the leptons (electrons and muons), we require a tight isolation criterion: no additional tracks with $p_T> 2$ GeV must be 
present in a cone $\Delta R = \sqrt{\Delta \eta^2 + \Delta \phi^2}= 0.5$ centred on the lepton track. The jets are reconstructed using only 
the calorimeter towers, through the anti-kt algorithm with cone size radius of 0.5, as defined in the {\tt FastJet} 
package~\cite{Cacciari:2011ma}  and implemented in {\tt Delphes}. The b-tagging efficiency is assumed to be 40\% for all b-jets, 
independently of their transverse momentum, with a fake rate of 1\% (10\%) for light (charm) jets. Finally, the total missing transverse 
energy (MET) is reconstructed using information from the calorimetric towers and muon candidates only.

In the following, we present a simple strategy that can lead to promising Signal-over-Background (S/B) ratios. Our purpose is to 
illustrate the new possibilities that open up in the \X\Xbar final state and motivate more detailed studies. To this aim detailed information 
on the efficiencies and the visible cross sections are given. 
\begin{table}[t!]
\begin{center}
\begin{tabular}{lcc}
\hline
 object & $|\eta_{max}|$ & $p_T^{min}$ [GeV]\\
\hline
e,$\mu$ & 2.4 & 10 \\ 
jets & 3 & 20 \\
b-jets & 2.5 & 20 \\
\hline
\end{tabular}  
\caption{Acceptance criteria of the various final states in the simulated detector.}
\label{Tab:DelphesAcceptance}
\end{center}
\end{table}

\section{Analysis and results}
As outlined before, we request two isolated same sign leptons in the final state. Two well reconstructed
leptons with $p_T>30$ GeV and $|\eta | < 2.4$ are sufficient to trigger the events (the two lepton combined trigger efficiency 
is assumed to be 100\% after the lepton selection cut). Selected jets have $p_T>$30 GeV and $|\eta |<$ 2.5. 
\begin{figure}
\centering
\epsfig{file=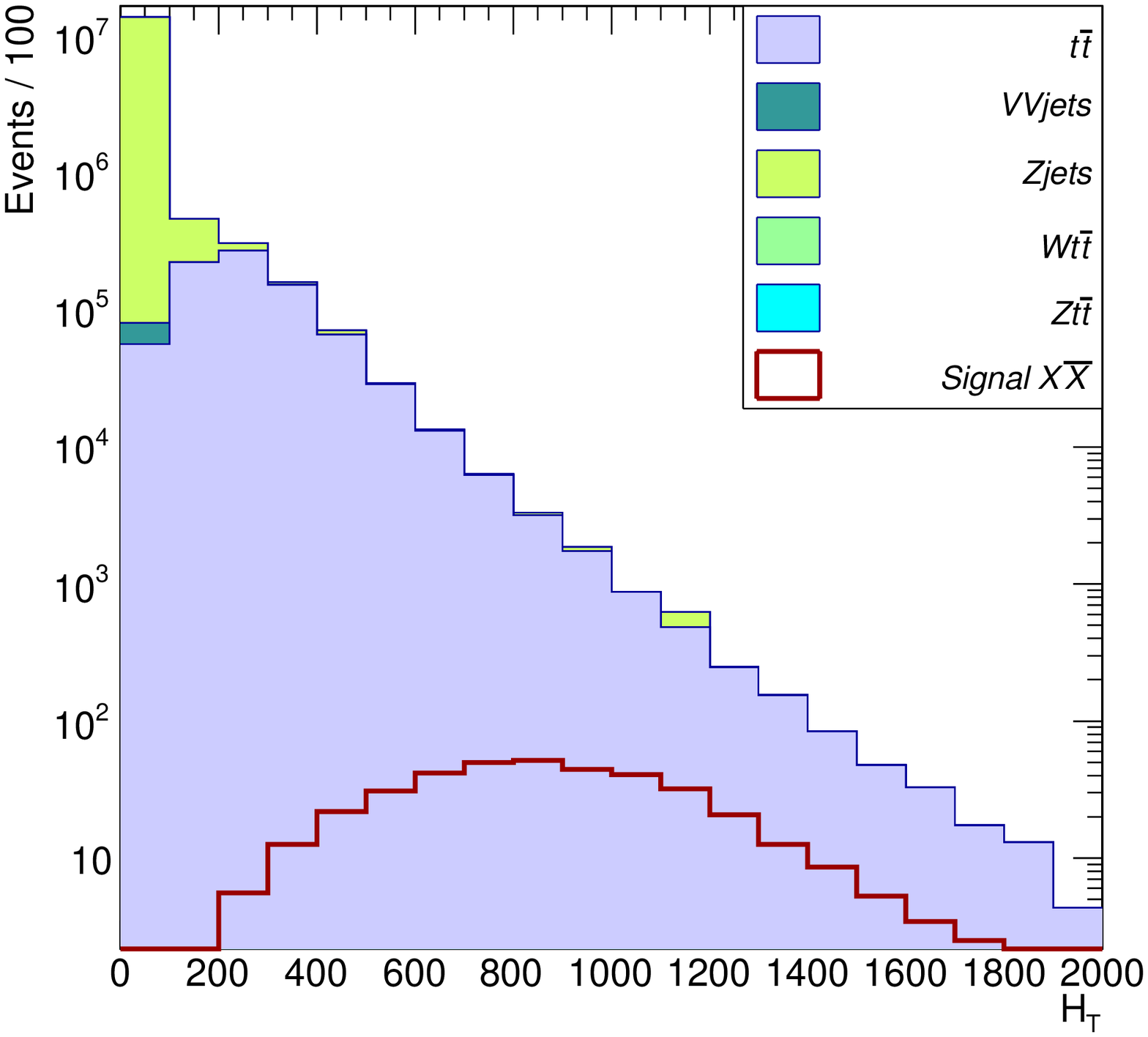, width=0.48\textwidth, angle=0}\hfill
\epsfig{file=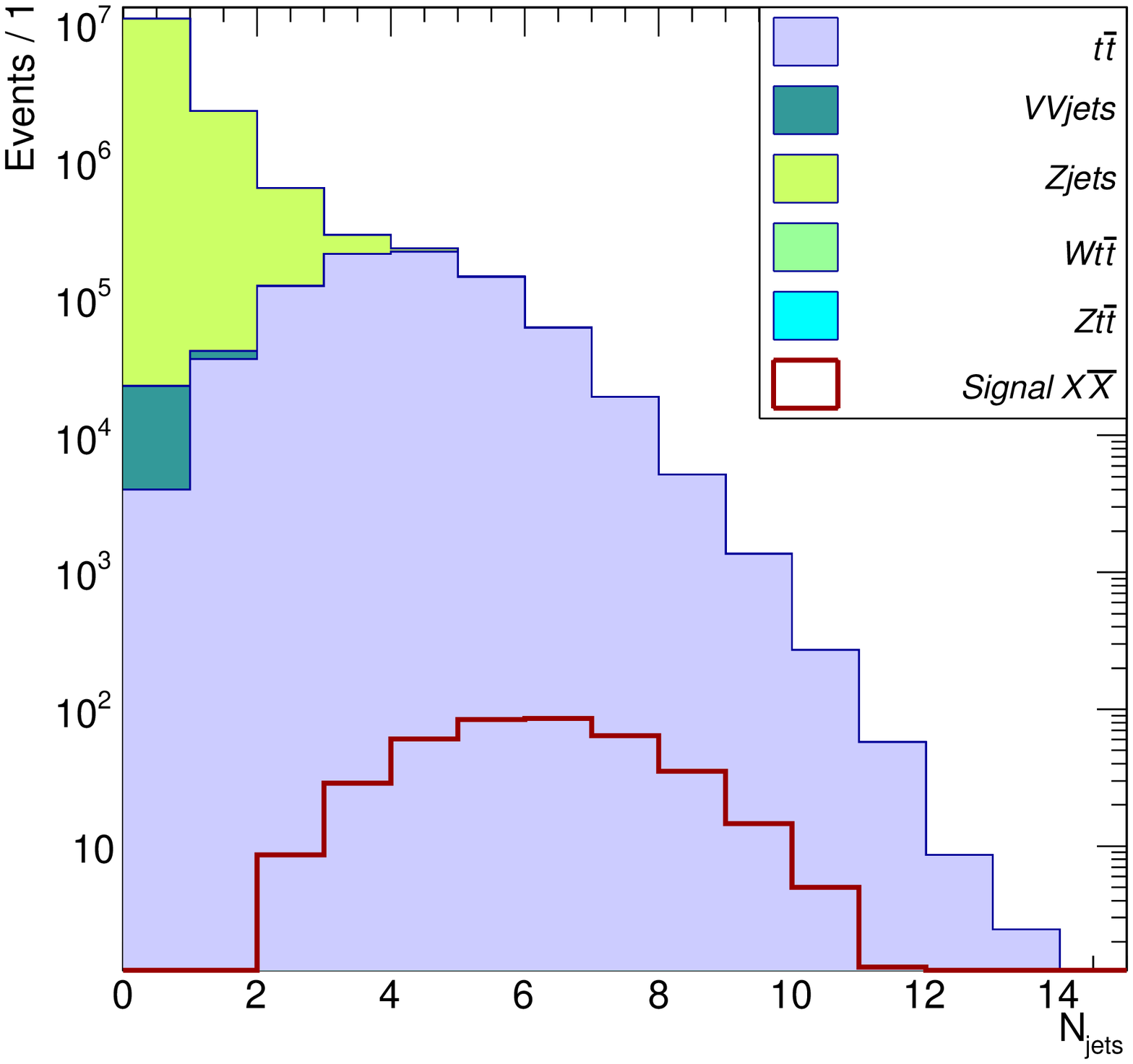, width=0.48\textwidth, angle=0}\hfill
\epsfig{file=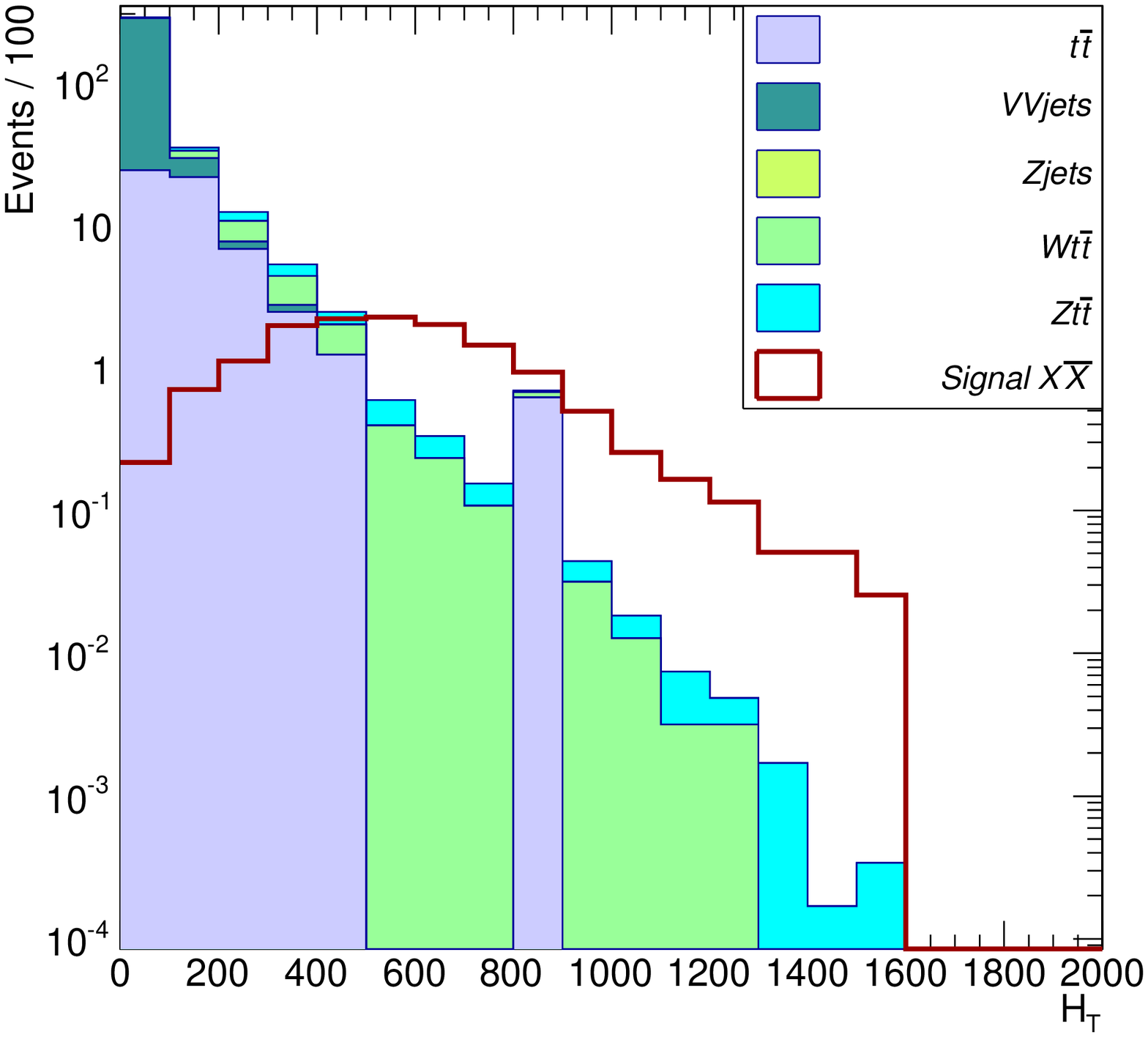, width=0.48\textwidth, angle=0}\hfill
\epsfig{file=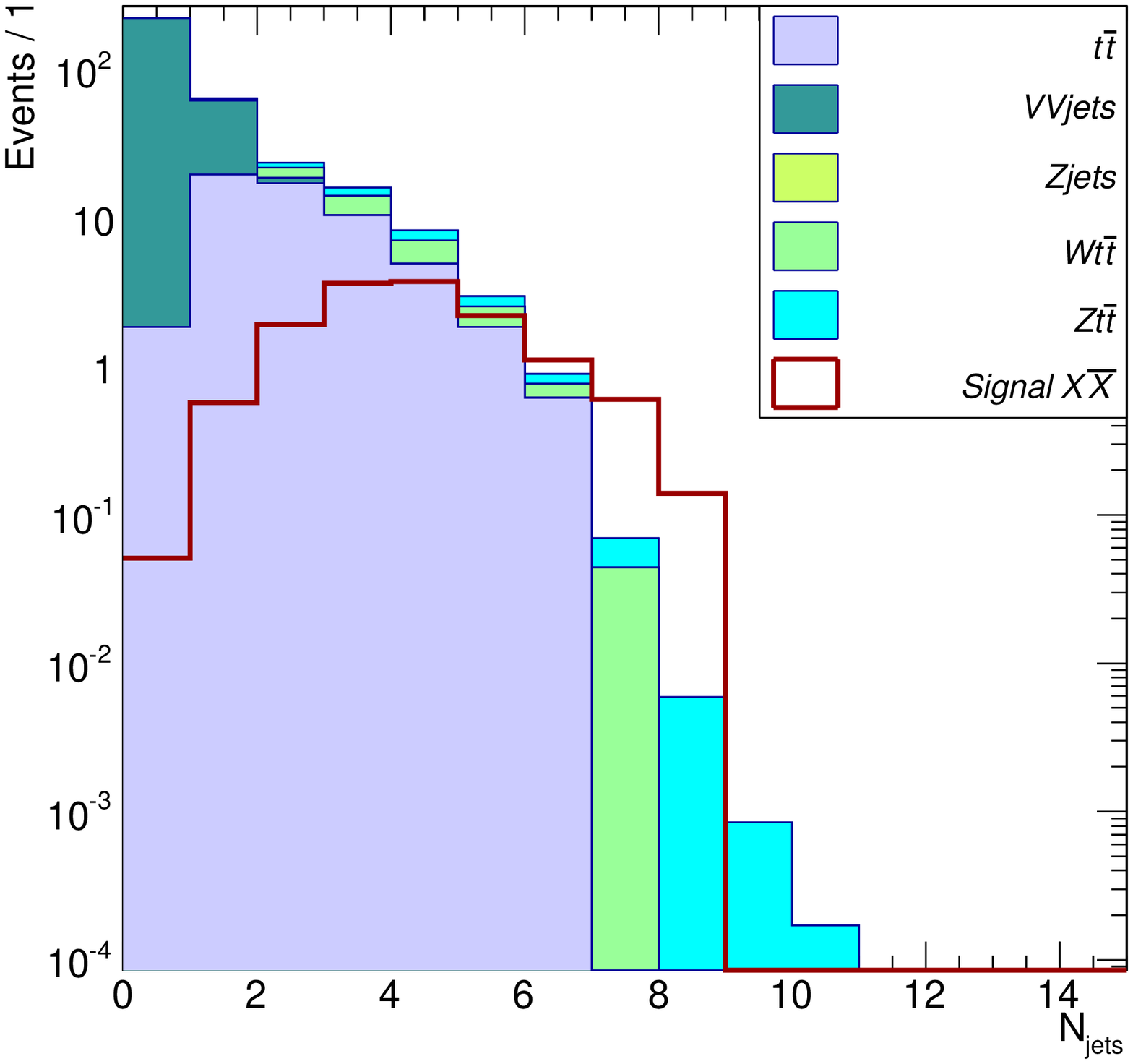, width=0.48\textwidth, angle=0}\hfill
\epsfig{file=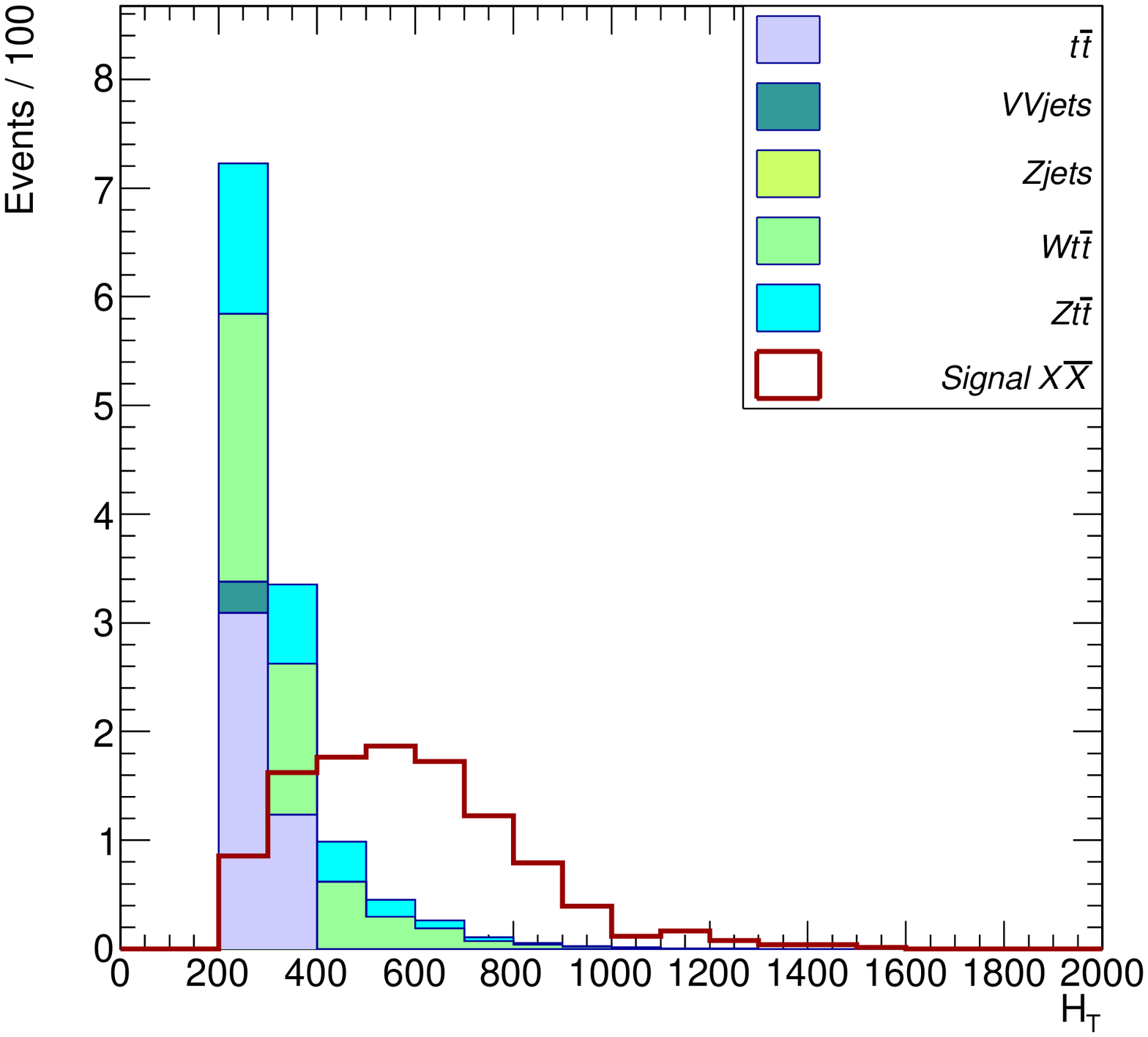, width=0.48\textwidth, angle=0}\hfill
\epsfig{file=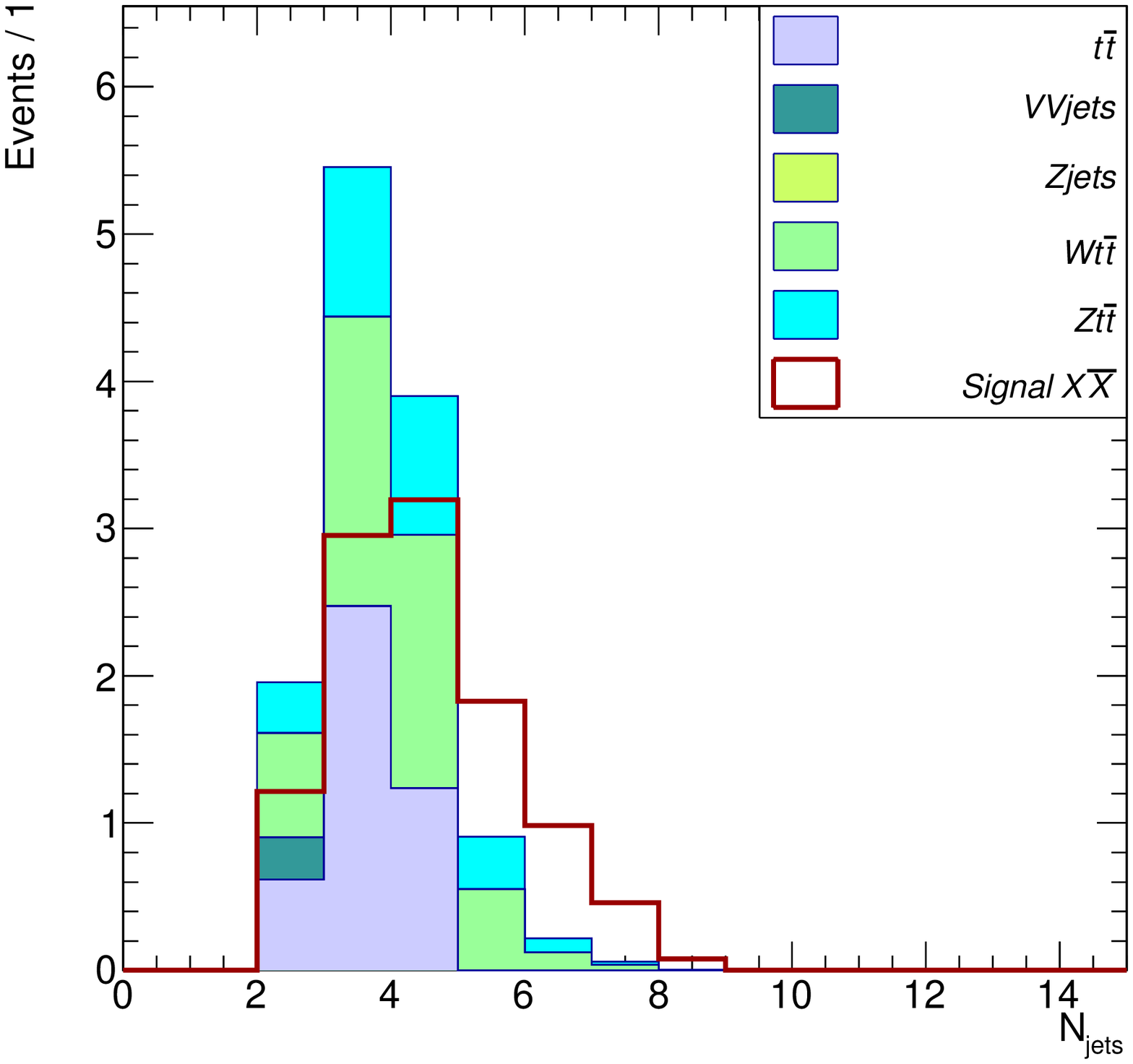, width=0.48\textwidth, angle=0}\hfill
{\caption{Number of expected events for a luminosity of 5fb$^{-1}$ for the signal and SM backgrounds (stacked histogram) with respect to the number of jets (left) and 
HT (right). The first row shows the distributions without selection, the middle row after the requirement of two same sign 
leptons, and the last row when requesting two same sign leptons, a $\Delta R_{min} (\ell,\mathrm{jet}) > 0.5$ and at least two jets.} 
\label{fig:dist}}
\end{figure}

The distributions of the number of jets and HT (which is defined as the scalar sum of all the selected jets in the event) after the 
various steps of the selection are shown on Figure~\ref{fig:dist} where for the signal the mass is 600 GeV/$c^2$. 
After requesting two same sign isolated leptons, the dominant background contributions are $VV$+jets and $t\bar{t}$. $VV$+jets can 
produce two real isolated leptons but signal and background are well separated both in the number of hard jets and HT. The $t\bar{t}$ 
contribution fulfils the two same sign isolated requirement when one lepton inside a b-jet can be seen as isolated. In order to help to 
remove this contribution, a cut on the angular distance between the lepton and the closest jet, $\Delta R_{min} (\ell,\mathrm{jet}) $, 
is used. The final selection, shown in the last row of Figure~\ref{fig:dist}, consists of two same sign isolated leptons, 
$HT \geq 200$ GeV, at least two jets and 
$\Delta R_{min} (\ell,\mathrm{jet}) > 0.5$. This selection removes most of the background and retains roughly all the signal events. For 
$M_X$=600 GeV/c$^2$ and an integrated luminosity of 5 fb$^{-1}$, the signal over background ratio is $3\cdot10^{-5}$ without selection, 
with the requirement of two same sign leptons it rises at $0.05$, and it is of the order of 1 for the final selection.
Because it has been obtained using a fast detector simulation and to be
distinguished from what is used in CMS, this selection will be referred to as {\it fast selection} in the following. 

\subsection{Mass limit for fixed BRs}
\label{boundsfixedBR}
\begin{figure}
\centering
\epsfig{file=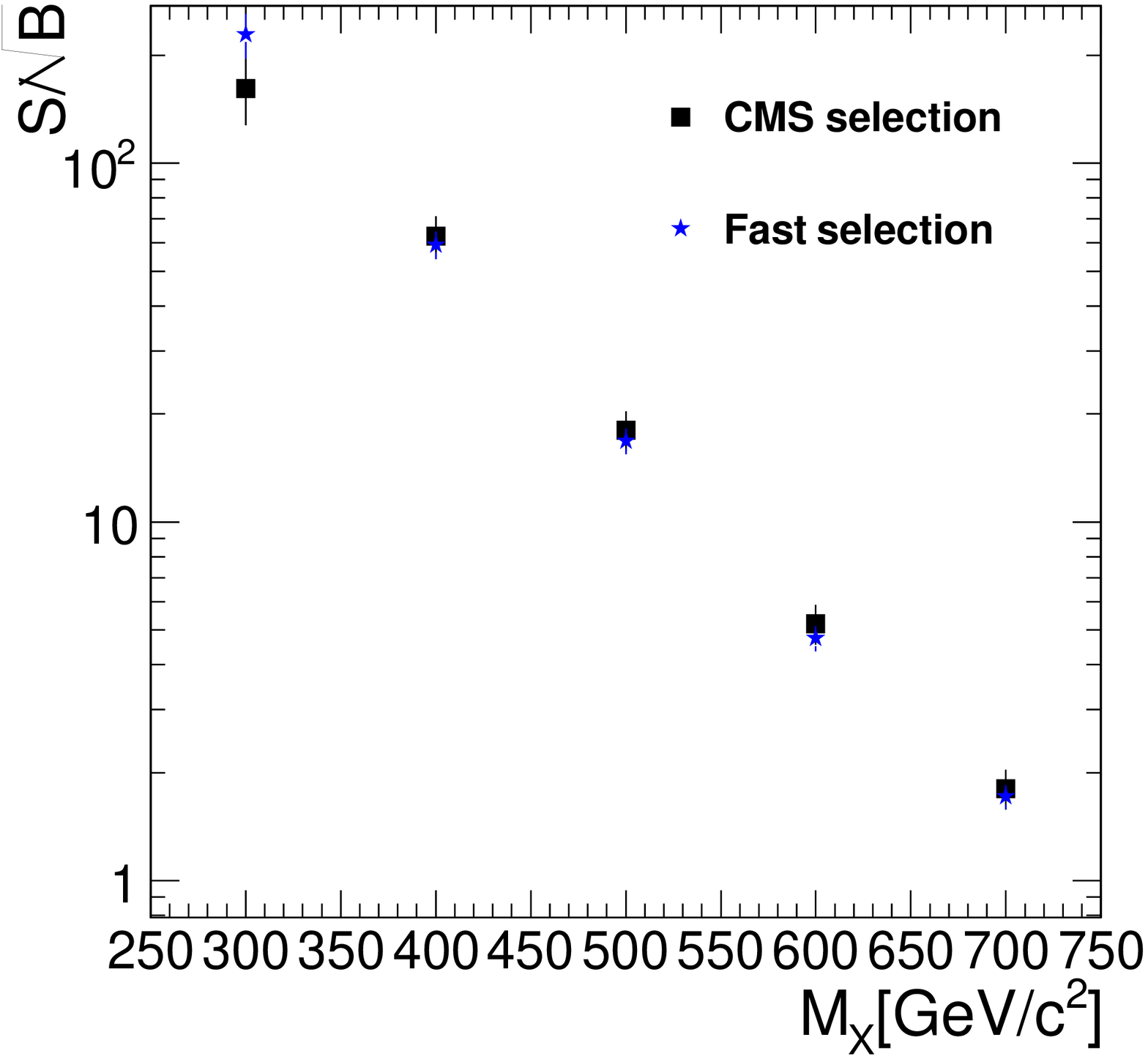, width=0.48\textwidth, angle=0}\hfill
\epsfig{file=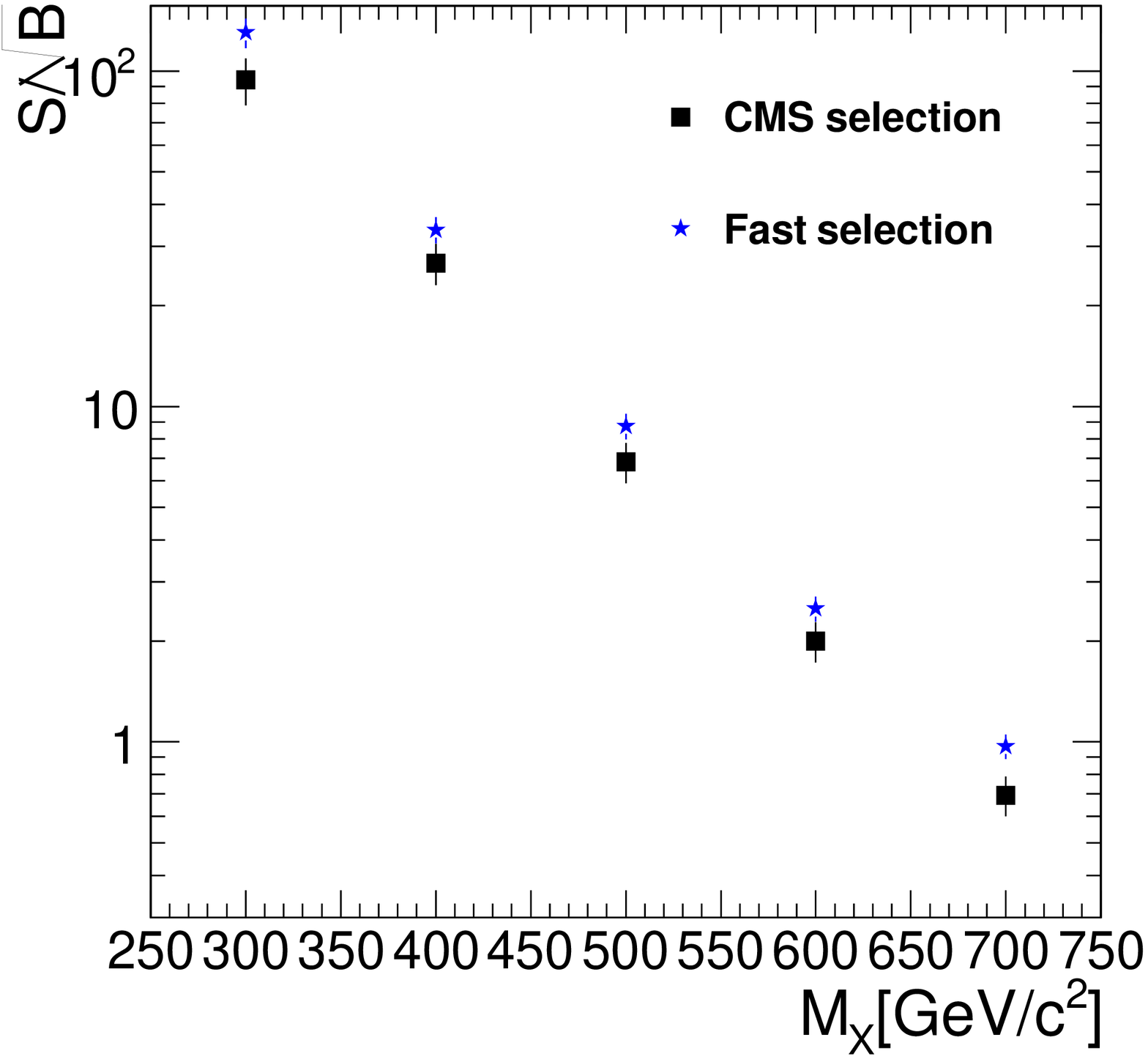, width=0.48\textwidth, angle=0}\hfill
{\caption{Significance of an eventual signal as a function of the
    signal generated mass, for fast and CMS selection, for events in
    which both \X decay via the Wt channel (left) and events in which
    one X decays to Wt and the other to Wc (right).} 
\label{fig:sigvsmass_XtXt_XtXc}}
\end{figure}

\begin{figure}
\centering
\epsfig{file=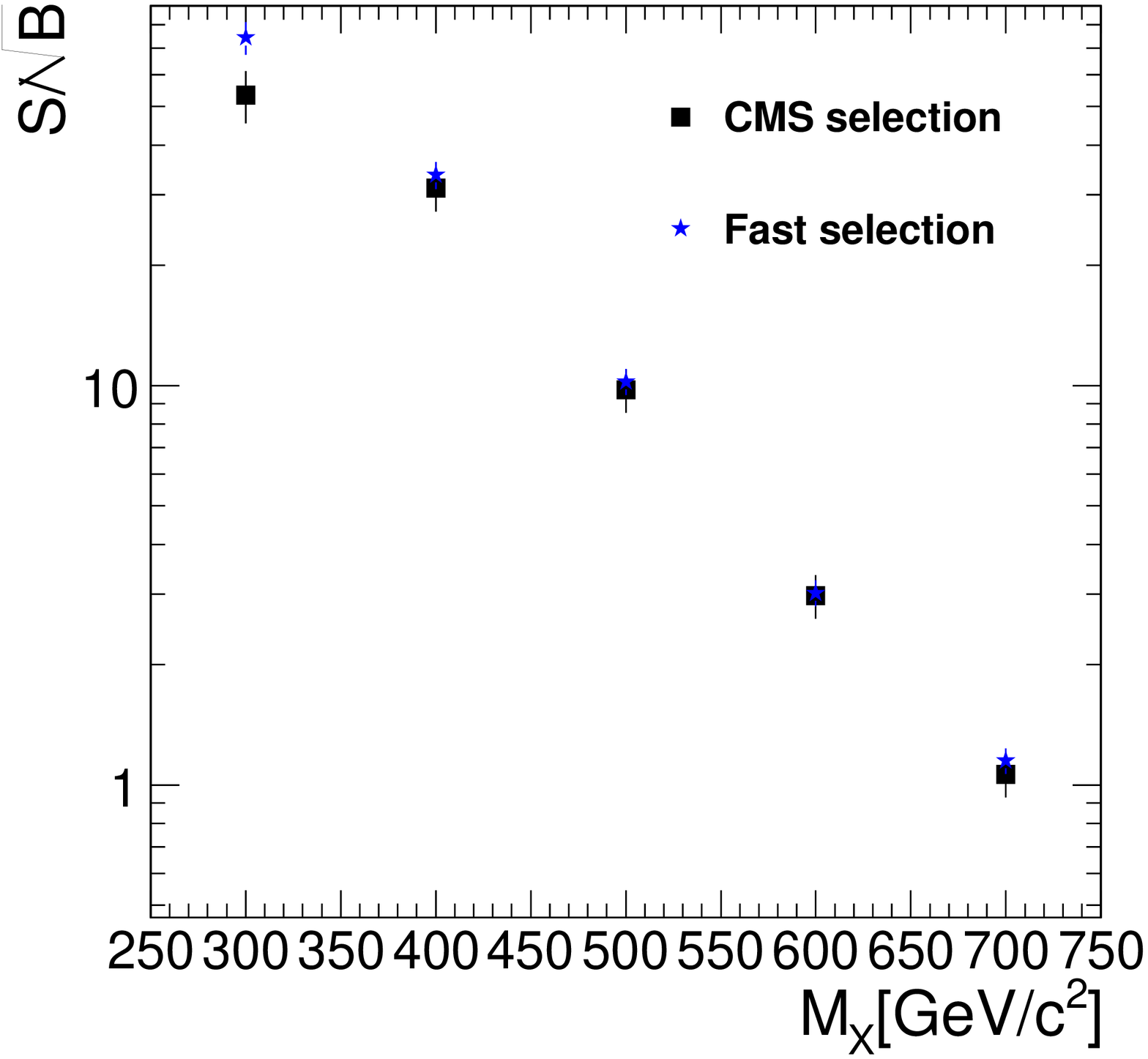, width=0.48\textwidth, angle=0}\hfill
{\caption{Significance of an eventual signal as a function of the signal generated mass, for fast and 
CMS selection, assuming $BR(X\to Wt)$ is fixed to a value depending on $M_X$.} 
\label{fig:sigvsmass_ALL}}
\end{figure}

The number of background (B) and signal (S) events surviving the final selection, is evaluated assuming an integrated 
luminosity of 5 fb$^{-1}$. The significance of an eventual signal is computed as $S/\sqrt{B}$.
Table~\ref{Tab:yield} and Figures~\ref{fig:sigvsmass_XtXt_XtXc} and \ref{fig:sigvsmass_ALL} show the evolution of the significance 
as a function of the mass.
In Table~\ref{Tab:yield} we also reported the efficiency on the inclusive signal after the same sign selection and after the final selection: the two numbers are quite close, thus showing that the additional cuts on $H_T$ and on the number of jets do not significantly affect the signal yield.
For the $ttH$ background we have used the cross section and branching ratios provided by the LHC Higgs Cross Section Working Group \cite{HXSWG}: considering a cross section at 7 TeV of 0.0863 pb, corresponding to a Higgs mass of 125 GeV, we have obtained a contribution to the background yield of (0.47$\pm$0.45), which is negligible with respect to the rest of the backgrounds and therefore will not be considered in the following analysis. 
This selection would allow an observation of a signal with a mass ranging from the tW production threshold up
to 561 GeV and an evidence up to 609 GeV.  These limits are valid assuming that the $BR(X\to Wt)$ is fixed to a value depending 
on $M_X$ in the specific model (the corresponding values are given in Table~\ref{channelweight}). 
\begin{table}[t!]
\begin{center}
\begin{tabular}{l|ccc|cc}
\hline
 Mass & Signal  & Background  & $S/\sqrt{B}$ & Efficiency & Efficiency\\
 (GeV)  & yield & yield   &   & final & same sign  \\
\hline
300 &  262.6 $\pm$ 17.3  & 12.5$\pm$1.66 & 74.3$\pm$6.9 & ($0.67\pm0.04$)\% & ($1.07\pm 0.055$)\% \\
400 &  119.1 $\pm$ 4.9  & 12.5$\pm$1.66 & 33.7$\pm$2.6 & ($1.73\pm0.07$)\% & ($2.28\pm0.08$)\%\\
500 &  36.2 $\pm$ 1.3  & 12.5$\pm$1.66 & 10.2$\pm$0.8  & ($2.24\pm0.08$)\% & ($2.86\pm0.09$)\% \\
600 &  10.7 $\pm$ 0.3  & 12.5$\pm$1.66 & 3.0$\pm$0.2 & ($2.36\pm0.08$)\% & ($3.08\pm0.09$)\%\\
700 &  4.1 $\pm$ 0.2 & 12.5$\pm$1.66 & 1.2$\pm$0.1 & ($2.84\pm0.09$)\% & ($3.52\pm0.095$)\%\\
\hline     
\end{tabular}  
\caption{Background and signal event yields for an integrated luminosity of $5$ fb$^{-1}$, after the final selection. The
$S/\sqrt{B}$ ratios is also provided. In the right-columns we also reported the inclusive efficiency after the final selection and after the same-sign dilepton selection.}
\label{Tab:yield}
\end{center}
\end{table}
Searches of this kind have been performed with LHC data. To compare with the actual available results, 
we reproduce the selection made in the CMS search \cite{CMSX53}.
This selection requires two well reconstructed leptons with $p_T>30$ GeV, at least four jets with $p_T>$30 GeV 
and $|\eta |<$ 2.5, $H_{T}>300$. The selection also includes a veto of the region around the Z mass peak. 
We use this alternative selection (which will be referred to as {\it CMS selection} in the following) as a reference. 
Since we are using a fast simulation, reproducing the reliability of the CMS study is behind our scope, nevertheless 
we want to check that we are able to reproduce reasonably the performances detailed in \cite{CMSX53}. 
Since the study presented in \cite{CMSX53} only considers the case in which $BR(X\to Wt)=1$, we 
restrict our study to such events in order to compare. 
The performance of such a selection on simulated signal events is found to be roughly comparable with the 
one described in \cite{CMSX53}. 

The significance of an eventual signal as a function of the signal generated mass, is shown in Figure~\ref{fig:sigvsmass_XtXt_XtXc} 
for the two selections, in the hypothesis of $BR(X\to Wt)=1$ (left plot). 
In this case the fast selection would allow an observation of a signal with a mass 
ranging from the tW production threshold up to 603 GeV and an evidence up to 653 GeV (when emulating the CMS
selection, the corresponding values are 609 and 653 GeV). 
In the right plot in Figure~\ref{fig:sigvsmass_XtXt_XtXc} we also show the significance of the signal if only the mixed decayed events are considered.
The signal selection efficiencies for different masses are listed in
Table \ref{Tab:Effs}, separately for events in which both \X decay via
the Wt channel (left) and events in which one \X decays to Wt and the
other to Wc (right). 
As the listed numbers include the effect of the $W$ boson branching ratio into leptons, in the case of equal efficiencies of the cuts, the efficiencies in the two channels should be $\epsilon$ (WtWt) = $2 \times \epsilon$ (WtWc): the numbers in the table, therefore, show that the {\it fast selection} is more efficient on the mixed decay sample than in the 100\% Wt sample.

\begin{table}[t!]
\begin{center}
\begin{tabular}{l|cc|cc|}
\hline
Mass & \multicolumn{2}{c|}{Final selection} &  \multicolumn{2}{c|}{Same-sign selection} \\
(GeV) & $\epsilon$ (WtWt) & $\epsilon$ (WtWc) & $\epsilon$ (WtWt) & $\epsilon$ (WtWc)  \\
\hline
300 & ($2.1\pm0.3$)\%  & ($1.2\pm0.1$)\% & ($3.7\pm 0.3$)\% & ($1.8 \pm 0.1$)\%\\
400 &  ($3.0\pm0.2$)\%  & ($1.7\pm0.1$)\% & ($4.1 \pm 0.2$)\% & ($2.3 \pm 0.1$)\%\\
500 &  ($3.7\pm0.2$)\%  & ($1.9\pm0.1$)\% & ($4.8 \pm 0.2$)\% & ($2.4 \pm 0.1$)\%\\
600 &  ($3.7\pm0.2$)\%  & ($1.9\pm0.1$)\% & ($4.8 \pm 0.2$)\% & ($2.6 \pm 0.1$)\% \\
700 &  ($4.2\pm0.2$)\% & ($2.4\pm0.1$)\% & ($5.4 \pm 0.2$)\% & ($2.9 \pm 0.1$)\%\\
\hline     
\end{tabular}  
\caption{Signal selection efficiencies for different masses listed separately for events in which both \X decay via
the Wt channel and events in which one \X decays to Wt and the other to Wc. In both cases, 
the efficiencies are calculated on the inclusive samples and not restricted to the same sign 
leptons sub-sample. For comparison, we also list the efficiencies after the same-sign selection.}
\label{Tab:Effs}
\end{center}
\end{table}

\begin{figure}
\centering
\epsfig{file=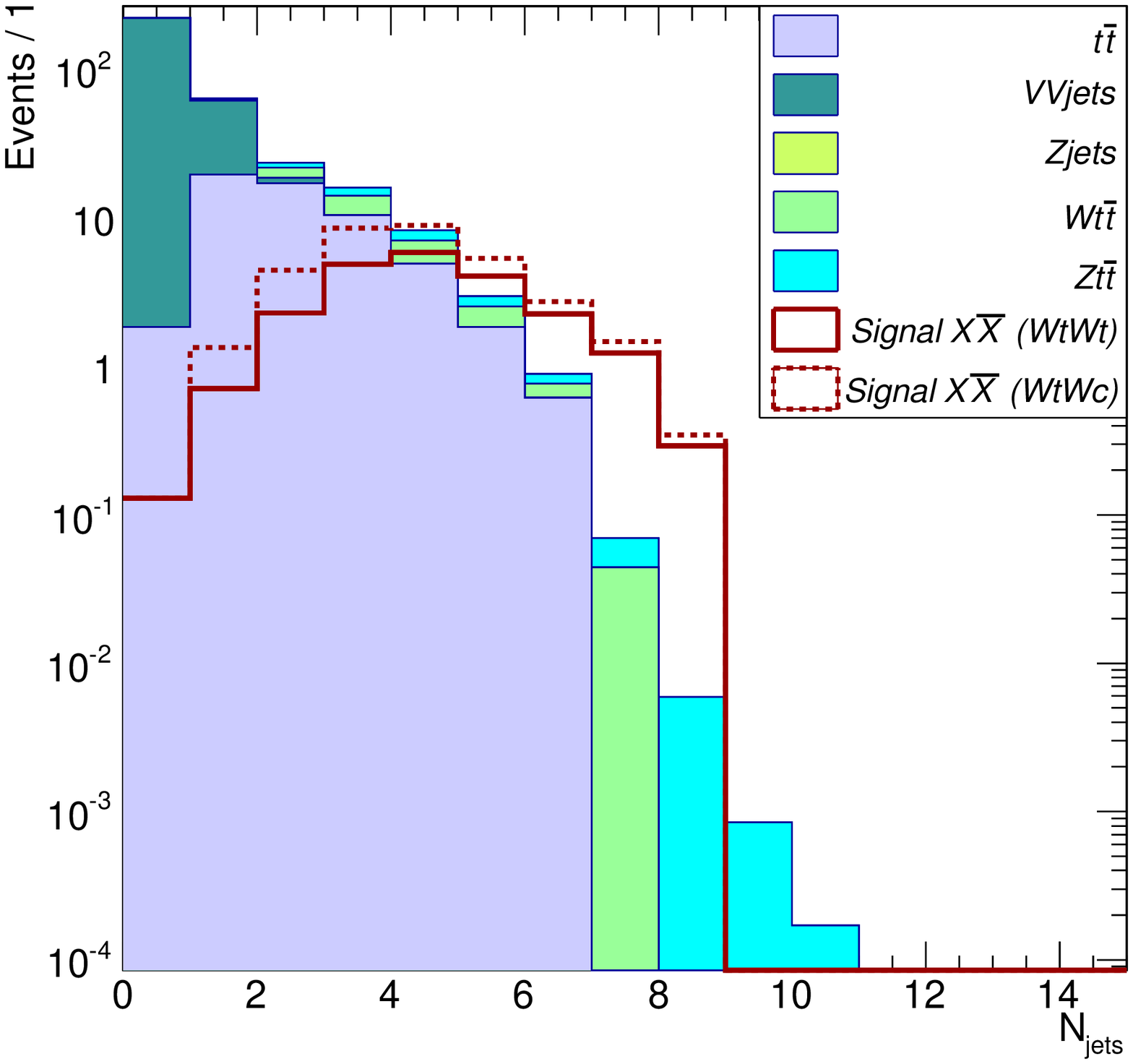, width=0.48\textwidth, angle=0}\hfill
\epsfig{file=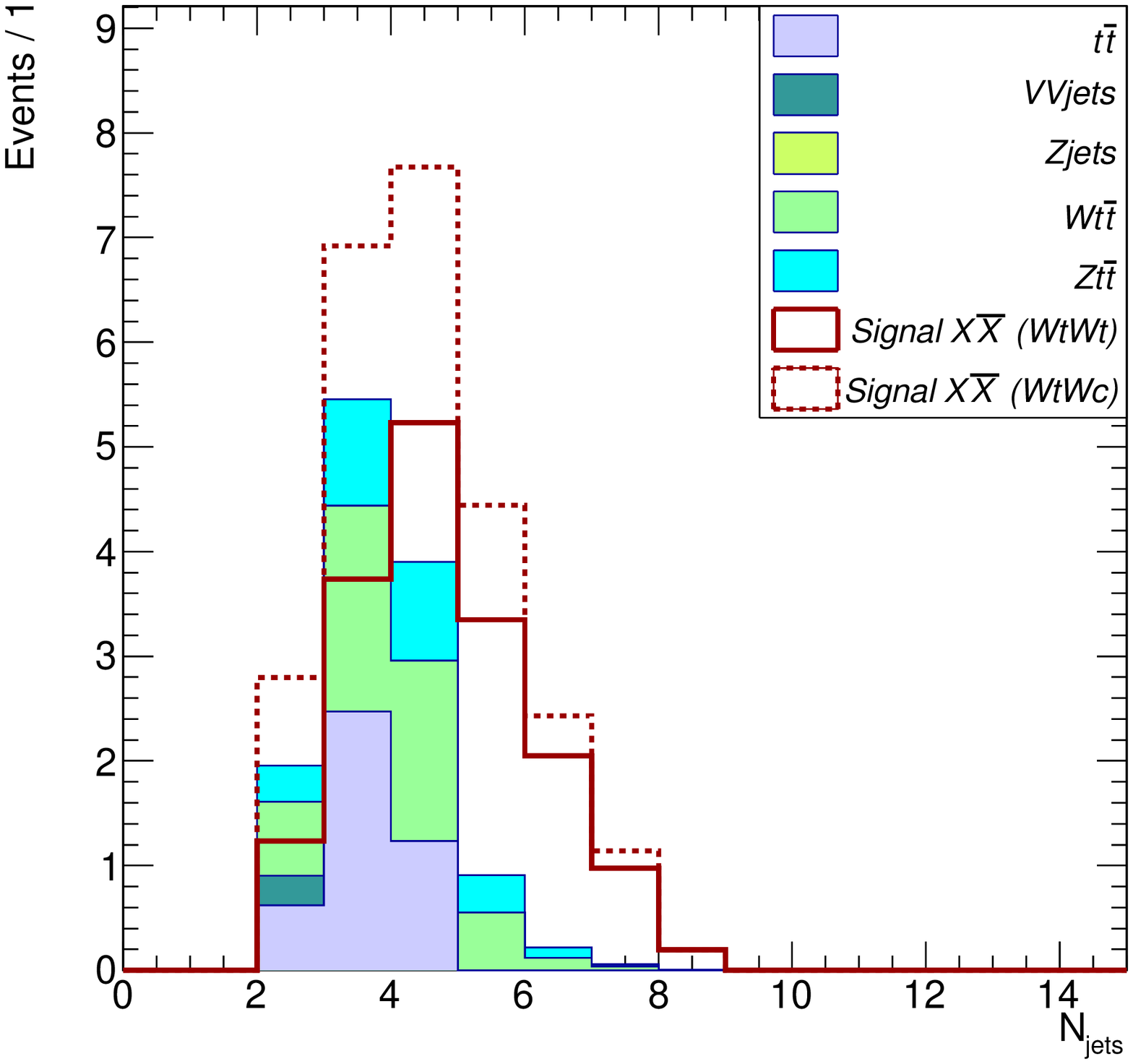, width=0.48\textwidth, angle=0}\hfill
{\caption{Number of expected events for a luminosity of 5fb$^{-1}$ for the signal and SM backgrounds (stacked histogram) with 
respect to the number of good jets in the event. The left plot shows the distributions after the requirement of two same sign 
leptons, and the right one when requesting two same sign leptons, a $\Delta R_{min} (\ell,\mathrm{jet}) > 0.5$ and at least two jets.} 
\label{fig:distNjets_tt_tc}}
\end{figure}

The same study has been repeated on signal samples for which the value of 
$BR(X\to Wt)$ is fixed to a value depending on $M_X$ (according to
Table~\ref{channelweight}), and hence composed of both events with
both \X decay via the Wt channel and events in which one \X decays to
Wt and the other one to Wc (the events where both \X decay via the Wc
channel are not surviving the request of two same sign leptons).
The CMS selection being tuned under the assumption of $BR(X\to Wt)=1$,
it is less optimal when one of the \X decays to Wc, while the
fast selection, tuned on a sample with $BR(X\to Wt)\neq 1$ (see exact 
values in Table~\ref{channelweight}) has better performances. 
As an example, the different cut on the number of jets in the event can have an impact, as can be seen in 
Figure~\ref{fig:distNjets_tt_tc}, where the distribution of the number of jets is shown after requesting two 
same sign leptons and after the final fast selection, separately for signal event where both \X decay in the  
Wt channel, and in the case where one \X decays to Wt and the other to Wc.
This comparison shows that considering the more general case $BR(X\to Wt)\neq1$ may lead to improved efficiency in the search for \X states.

The reach of this analysis is further enhanced in the recent data collected by the LHC at 8TeV.
From a rough evaluation using the efficiencies obtained in the present study, taking into account
the increase in cross sections (for both signal and background processes), and assuming an
integrated luminosity of 19/fb (corresponding to the data collected by CMS at 8TeV in 2012), we expect the significance
to increase by a factor of 2.5--3 for the different mass points considered.
This selection would thus allow to raise the bound on \X mass to above 700 GeV for having an evidence of the signal.

\subsection{Analysis with BR($X_{5/3}\to tW$) as a free parameter}
\label{freeBR}
\begin{figure}
\centering
\epsfig{file=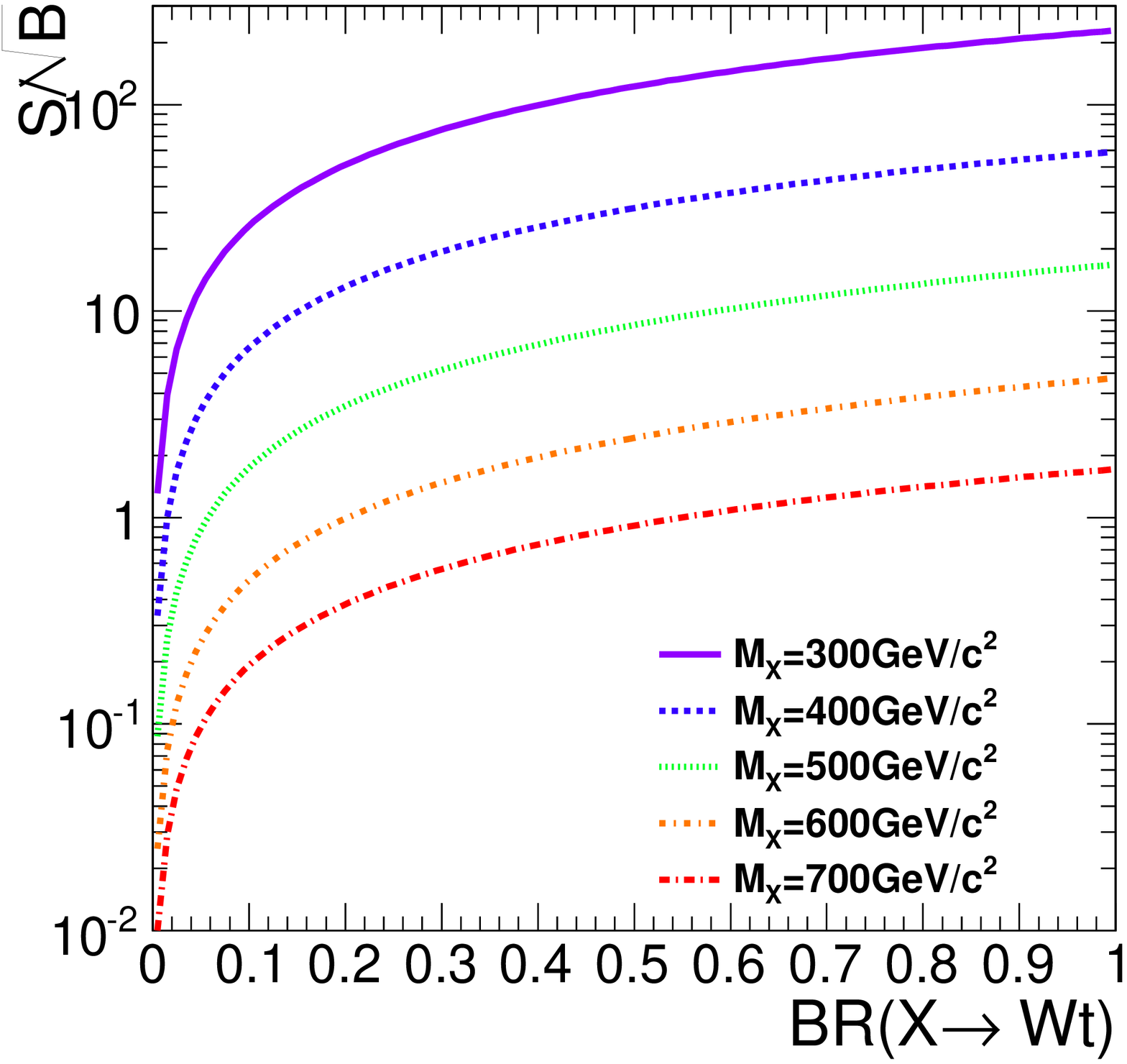,width=.48\textwidth} \epsfig{file=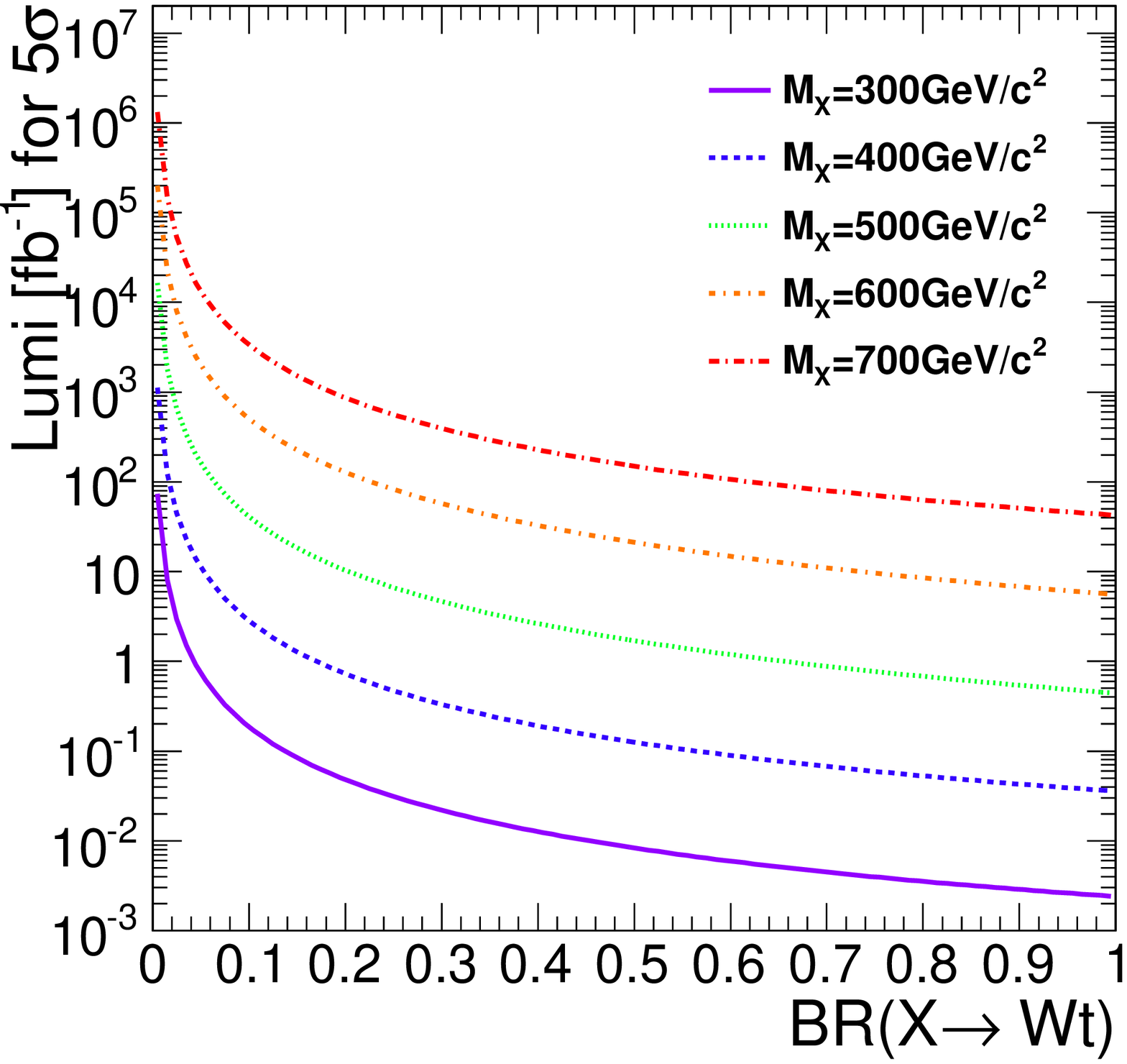,width=.48\textwidth}
\caption{Significance for an integrated luminosity of 5/fb and luminosity needed for a $5\sigma$ significance as function of $BR(X\to Wt)$ for different $M_X$ values.} 
\label{fig:SBvsRB}
\end{figure}

\begin{figure}
\centering
\epsfig{file=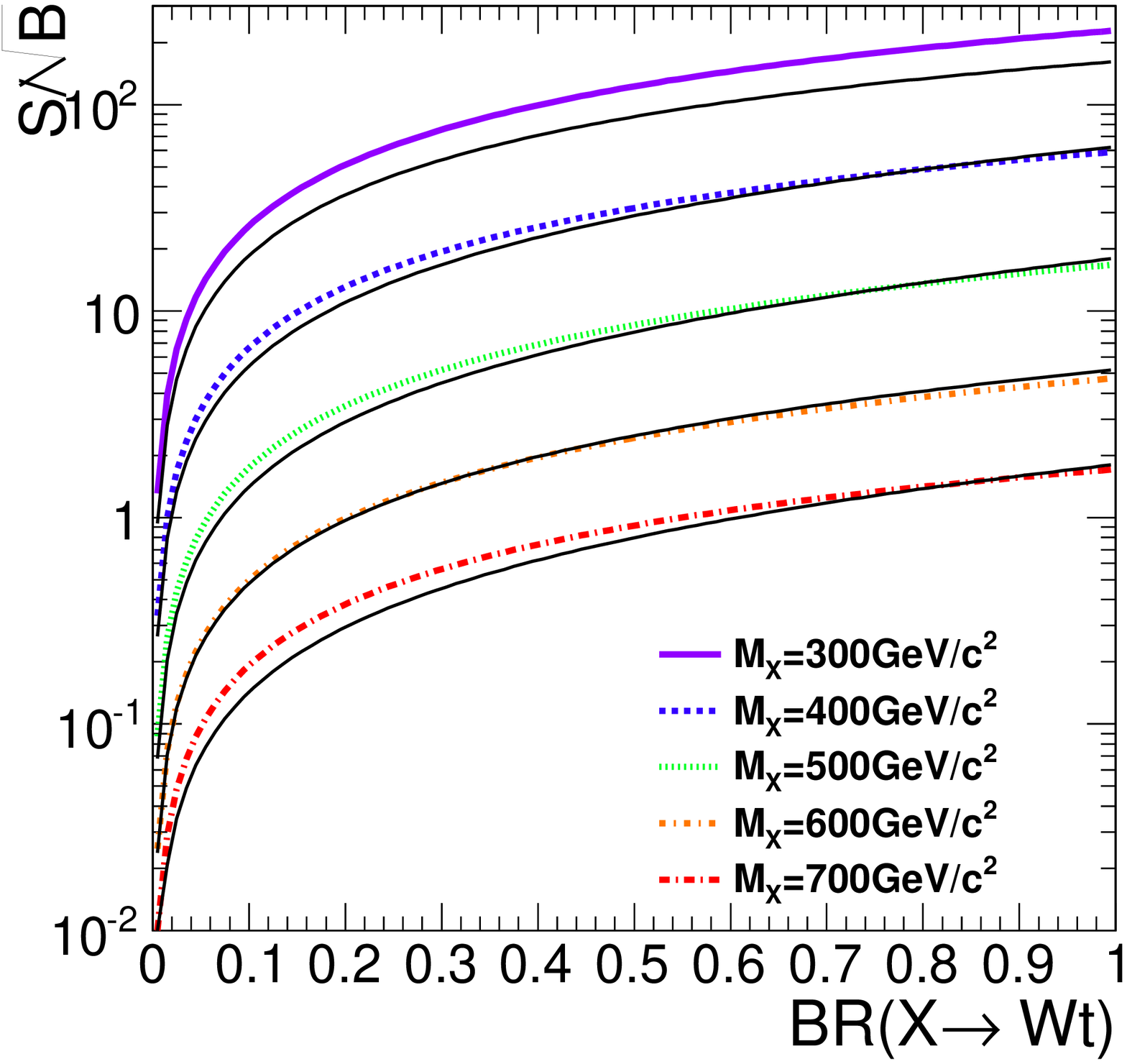,width=.5\textwidth}
\caption{Significance as function of $BR(X\to Wt)$  for different $M_X$ values for an integrated luminosity of 5/fb, compared 
for fast selection (coloured thick lines) and CMS selection (solid narrow lines).} 
\label{fig:SBvsRB_DRvsCMS}
\end{figure}

In the previous sections, the magnitude of the decay rate of \X to Wt, $BR(X\to Wt)$ was chosen depending on the value of $M_X$ 
(see Table~\ref{channelweight}). In order to have an idea of the possible reach and limits without imposing a specific link between 
mass and branching ratios,
in this section we let the decay rate free to vary between 0 and 1. We study the significance of such a signal as a function of  the 
decay rate. The study is performed separately and independently for each mass point, and we assume $BR(X\to Wt)=1-BR(X\to Wc)$
(that is $V_R^{42}$ maximal; we discussed in Sect.~\ref{sec:proddecay} that the particular choice of mixing in the light sector does
not change significantly the production cross section). Figure~\ref{fig:SBvsRB} shows the significance $S/\sqrt{B}$ as a function of 
$BR(X\to Wt)$  for different $M_X$ values. As expected the significance increases with $BR(X\to Wt)$. 
For masses above 600 GeV/c$^2$, it is not possible to obtain an evidence of the signal with an integrated luminosity of 5/fb.
In the right panel of Figure~\ref{fig:SBvsRB} we also show the needed luminosity for a 5$\sigma$ significance as a function of $BR(X\to Wt)$  for different 
$M_X$ values. For a mass of 600 GeV/c$^2$, an integrated luminosity of 6/fb is needed in the best case scenario. 
In Figure~\ref{fig:SBvsRB_DRvsCMS}, the evolution of the significance with the value of $BR(X\to Wt)$ is shown 
for the fast selection, compared to the reference CMS selection.
The plot shows that significant improvements may be achieved for comparable decay rates into tops and light quarks.
The results of this study are only qualitative, as a more realistic simulation of the detector performance and optimisation of the cut for various masses would be required to validate it.

\section{Conclusions}
The search for a vector-like quark \X gives important bounds on the structure of possible extensions of the SM. In this paper we 
have used the requirement for two isolated leptons of the same sign to analyse the signal coming from pair production of the 
exotic \X vector-like quark at LHC. We have performed a detailed analysis using a Monte Carlo simulation taking into account both 
signal and background and a simplified detector simulation to have a more detailed idea of the perspectives for detection or 
exclusion for the LHC experiments. 
The novelty of our study is that we consider the most general scenario, where the exotic \X quark is allowed to decay both into $W^+ t$ and in light quarks, $W^+ c$ and $W^+ u$.
We point out that the channel with one \X decaying into a top and the other into a light quark will also contribute to the same sign 
dilepton signal, and that it is crucial to know the efficiency of the experimental search on this channel, together with the 
$W^+ t W^- \bar{t}$ in order to extract a realistic bound on the \X mass.
We also point out the possible relevance of single \X production.

After validating our simulation with the dedicated CMS search, where 100\% decays into tops are assumed, we propose a different set of cuts, with a lower cut in $H_T$, a minimal cut in the number of jets and using the lepton isolation as the main cut to reduce the $t \bar{t}$ background.
Our results show that the two sets of cuts give similar significance $S/\sqrt{B}$ in the 100\% $X_{5/3} \to W^+ t$ case, while our selection offers a better efficiency in the mixed case.
In the case of large rate into light quarks, the same sign dilepton strategy is complemented by the CMS search where heavy quarks decaying into $W$ plus light jets are fully reconstructed.
While we do not advocate our selection cuts as being optimal, the results of our study motivate a more detailed study of the LHC discovery potential of a vector-like quark with charge $+5/3$, beyond the approximation of 100\% decays into tops.

\section*{Acknowledgements}
This project was partially supported by a ``Theory LHC France 2012'' grant from IN2P3. We also acknowledge the warm hospitality 
of the CMS centre in Lyon for discussions.

\end{document}